\documentclass[a4paper,12pt]{article}
\pdfoutput=1
\usepackage[utf8]{inputenc}
\usepackage{amsmath,amssymb}
\usepackage{bm} 
\usepackage{graphicx}
\usepackage{pdflscape}
\usepackage{tabularx}
\usepackage{setspace}
\usepackage[margin=0.75in]{geometry}
\usepackage[skip=2pt]{caption}
\newcolumntype{S}{>{\centering\arraybackslash}m{0.08\textwidth}}
\newcolumntype{M}{>{\centering\arraybackslash}m{0.15\textwidth}}
\newcolumntype{L}{>{\centering\arraybackslash}m{0.2\textwidth}}
\usepackage[table]{xcolor}

\usepackage[hypertexnames=false]{hyperref}
\usepackage{float}

\usepackage{natbib} 
\setcitestyle{square,aysep={},yysep={;}}
\DeclareCaptionFormat{myformat}{\fontsize{10}{12}\selectfont#1#2#3}
\title{Predicting non-linear dynamics by stable local learning in a recurrent spiking neural network}
\author{Aditya Gilra$^{1\textnormal{,*}}$ \and Wulfram Gerstner$^1$}
\date{%
\begin{singlespace}\begin{normalsize}
    $^1$School of Computer and Communication Sciences, and Brain-Mind Institute, School of Life Sciences, \'{E}cole Polytechnique F\'{e}d\'{e}rale de Lausanne (EPFL), Lausanne 1015, Switzerland.\\%
    $^*$Correspondence: aditya.gilra@epfl.ch\\[2ex]
\end{normalsize}\end{singlespace}
}

\begin{document}

\maketitle

\begin{abstract}
Brains need to predict how the body reacts to motor commands. It is an open question how networks of spiking neurons can learn to reproduce the non-linear body dynamics caused by motor commands, using local, online and stable learning rules. Here, we present a supervised learning scheme for the feedforward and recurrent connections in a network of heterogeneous spiking neurons. The error in the output is fed back through fixed random connections with a negative gain, causing the network to follow the desired dynamics, while an online and local rule changes the weights. The rule for Feedback-based Online Local Learning Of Weights (FOLLOW) is local in the sense that weight changes depend on the presynaptic activity and the error signal projected onto the postsynaptic neuron. We provide examples of learning linear, non-linear and chaotic dynamics, as well as the dynamics of a two-link arm. Using the Lyapunov method, and under reasonable assumptions and approximations, we show that FOLLOW learning is stable uniformly, with the error going to zero asymptotically.
\end{abstract}

\section{Introduction}

\begin{figure}
\begin{center}
\includegraphics[width=0.95\textwidth]{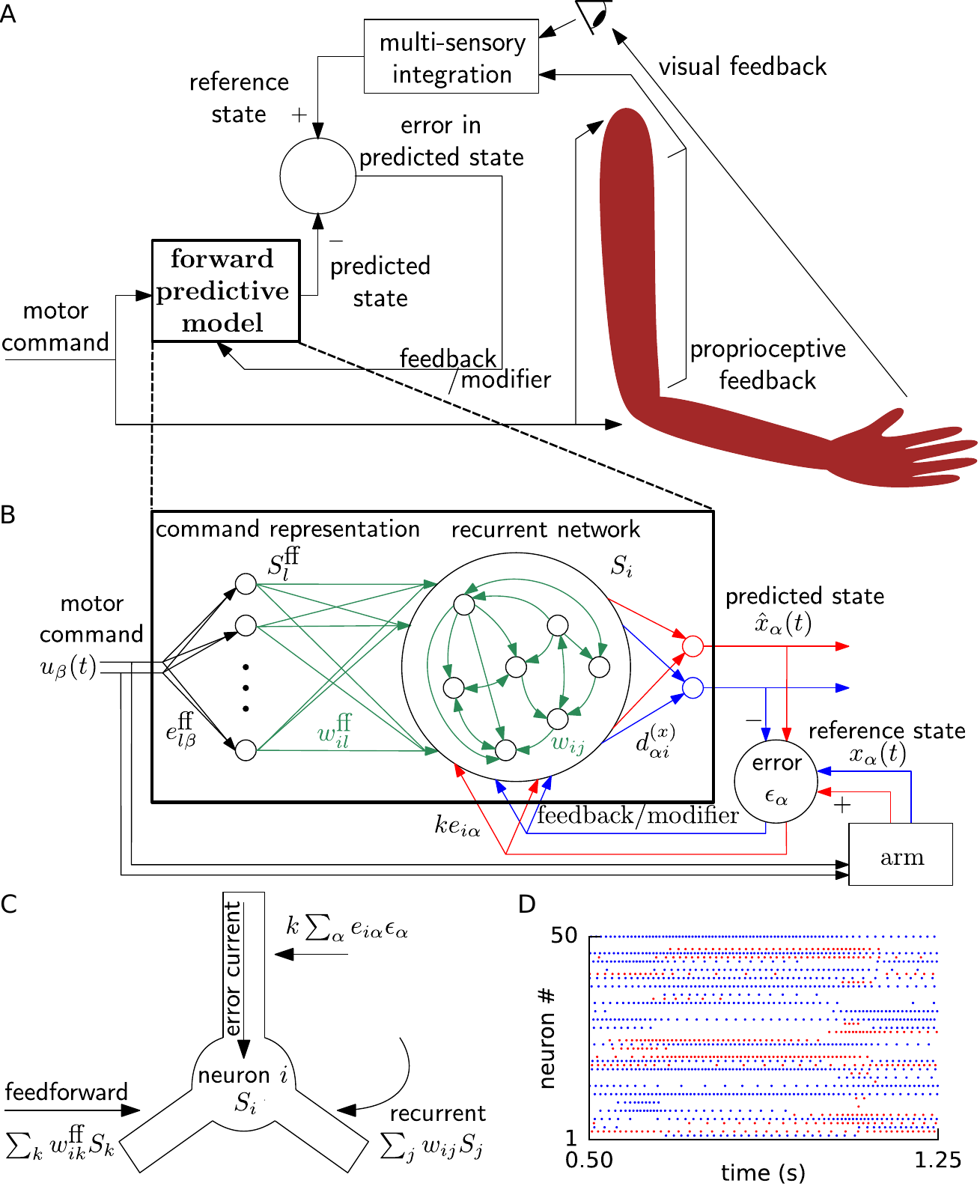}
\end{center}
\caption{\textbf{Schematic for learning a forward model} \textbf{A.} During learning, random motor commands (motor babbling) cause movements of the arm, and are also sent to the forward predictive model, which must learn to predict the positions and velocities (state variables) of the arm. The deviation of the predicted state from the reference state, obtained by visual and proprioceptive feedback, is used to learn the forward predictive model with architecture shown in B.
\textbf{B.} Command input $\vec{u}$ projected onto neurons with random weights $e^{\textnormal{ff}}_{k\alpha}$. The spike trains of these command representation neurons $S_l^\textnormal{ff}$ are sent via plastic feedforward weights $w^{\textnormal{ff}}_{il}$ into the neurons of the recurrent network having plastic weights $w_{ij}$ (plastic weights in green). Readout weights $d_{\alpha i}$ decode the filtered spiking activity of the recurrent network as the predicted state $\hat{x}_\alpha(t)$. The deviations of the predicted state from the reference state is fed back into the recurrent network (red and blue for different values of index $\alpha$) with encoding weights $ke_{i\alpha}$. \textbf{C.} A cartoon depiction of feedforward, recurrent and error currents entering a neuron $i$ in the recurrent network. The error current enters the apical dendrite and triggers intra-cellular signals available at the synapses for weight updates, isolated from the somatic current (here shown via basal dendrites). \textbf{D.} A few spike trains of neurons of the recurrent network from the non-linear oscillator example are plotted.}
\label{fig:schematics}
\end{figure}

\paragraph{}
Over the course of life, we learn many motor tasks such as holding a pen, chopping vegetables, riding a bike or playing tennis. To control and plan such movements, the brain must implicitly or explicitly learn forward  models \citep{conant_every_1970} that predict how our body responds to neural activity in brain areas known to be involved in  motor control (Figure \ref{fig:schematics}A). More precisely, the brain must acquire a representation of  the dynamical system formed by our muscles, our body, and the outside world in a format that can be used to plan movements and initiate corrective actions if the desired motor output is not achieved \citep{pouget_computational_2000,wolpert_computational_2000,lalazar_neural_2008}.
Visual and / or  proprioceptive feedback from spontaneous movements during 
pre-natal \citep{khazipov_early_2004} and post-natal development \citep{petersson_spontaneous_2003} or from voluntary movements during adulthood \citep{wong_can_2012,hilber_motor_2001} are important to learn how the body moves in response to neural motor commands
\citep{lalazar_neural_2008,wong_can_2012,sarlegna_roles_2009,dadarlat_learning-based_2015}, and how the world reacts to these movements \citep{davidson_widespread_2005,zago_fast_2005,zago_visuo-motor_2009,friston_hierarchical_2008}. We wondered whether a non-linear dynamical system, such as a forward predictive model of a simplified arm, can be learned and represented in a heterogeneous network of spiking neurons by adjusting the weights of recurrent connections.


\paragraph{}
Supervised learning of recurrent weights to predict or generate non-linear dynamics, given command input, is known to be difficult in networks of rate units, and even more so in networks of spiking neurons \citep{abbott_building_2016}.
Ideally, in order to be biologically plausible, a learning rule must be \textit{online} i.e. constantly incorporating new data, as opposed to batch learning where weights are adjusted only after many examples have been seen; and \textit{local} i.e. the quantities that modify the weight of a synapse must be available locally at the synapse as opposed to
backpropagation through time (BPTT) \citep{rumelhart_learning_1986} or
real-time recurrent learning (RTRL) \citep{williams_learning_1989} which are  non-local in time or in space, respectively \citep{pearlmutter_gradient_1995,jaeger_tutorial_2005}.
Even though Long-Short-Term-Memory (LSTM) units \citep{hochreiter_long_1997}
avoid the vanishing gradient problem 
\citep{bengio_learning_1994,hochreiter_gradient_2001} in recurrent networks, the corresponding learning rules are difficult to interpret biologically.

\paragraph{}
Our approach toward learning of recurrent spiking networks is situated at the crossroads of reservoir computing \citep{jaeger_echo_2001,maass_real-time_2002,legenstein_input_2003,maass_computational_2004,jaeger_harnessing_2004,joshi_movement_2005,legenstein_edge_2007}, FORCE learning \citep{sussillo_generating_2009,sussillo_transferring_2012,depasquale_using_2016,thalmeier_learning_2016,nicola_supervised_2016}, and adaptive control theory \citep{morse_global_1980,narendra_stable_1980,slotine_adaptive_1986,slotine_adaptive_1987,narendra_stable_1989,sastry_adaptive_1989,ioannou_robust_2012}.
In contrast to the original reservoir scheme  \citep{jaeger_echo_2001,maass_real-time_2002} or neural network implementations of control theory
\citep{sanner_gaussian_1992,dewolf_spiking_2016} where learning was restricted to the read-out or feedforward connections, we focus on a learning rule for the recurrent connections; and in contrast to FORCE learning where recurrent synaptic weights have to change rapidly during the initial phase of learning \citep{sussillo_generating_2009,sussillo_transferring_2012}, we aim for a learning rule that works in the biologically more plausible setting of slow synaptic changes. While previous work has shown that linear dynamical systems can  be represented and  learned with local online rules in recurrent spiking networks \citep{macneil_fine-tuning_2011,bourdoukan_enforcing_2015}, for  non-linear dynamical systems the recurrent weights in spiking networks have typically been computed offline \citep{eliasmith_unified_2005}.

\paragraph{}
Here, we propose a scheme for how a recurrently connected network of heterogeneous deterministic spiking neurons may learn to mimic a low-dimensional non-linear dynamical system, with a local and online learning rule.
The proposed learning rule is supervised, and requires access to the error in observable outputs. The output errors are fed back with random, but fixed feedback weights. Given a set of fixed error-feedback weights, the learning rule is synaptically local and combines presynaptic activity with the local postsynaptic error variable.

\section{Results}

\paragraph{}
A forward predictive model (Fig. \ref{fig:schematics}A) takes, at each time step,
a motor command $\vec{u}(t)$ as input and predicts the next observable state $\vec{\hat{x}}(t+\Delta t)$ of the system.
In the numerical implementation, we consider $\Delta t = 1$ms, but for the sake of notational simplicity we drop the $\Delta t$ in the following.
The predicted system state $\vec{\hat{x}}$ (e.g., the position and velocity of the hand) is assumed to be low-dimensional with dimensionality $N_d$ (4-dimensional for a two-link arm).
The motor command $\vec{u}(t)$ is used to generate target movements such as ``lift your arm to a location'', with its dimensionality $N_c$ typically smaller than the dimensionality $N_d$ of the system state.

\paragraph{}
In  our  neural network model, the low dimensional motor command drives the spiking activity of a command representation layer of 3,000 to 5,000 integrate-and-fire neurons (Fig. \ref{fig:schematics}B) via connections with fixed random weights.
These neurons project, via plastic feedforward connections, to a recurrent network of also 3000 to 5000 integrate-and-fire neurons. We assume that the predicted state $\hat{x}$ is linearly decoded from the activity of the recurrent network.
If we denote the spike train of neuron $i$ by $S_i(t)$, then component $\alpha$ of the predicted system state is
$\hat{x}_\alpha(t)  = \sum_i d_{\alpha i} \int_{-\infty}^t S_i(s)\kappa(t-s)ds$
where $d_{\alpha i}$ are the readout weights. The integral represents a convolution with a low-pass filter $\kappa(t)$ with a time constant of 20 ms,
and will be denoted by $(S*\kappa)(t)$ in the following.

\paragraph{}
The predicted output is compared with the observable system state $x(t)$, e.g.,
the position and velocity of the hand deduced from visual and proprioceptive input.
A nonlinear dynamical control system will evolve generally as $d\vec{x}/dt = \vec{h}(\vec{x}(t),\vec{u}(t))$, but we first consider a simpler dynamical system given by a set of coupled differential equations
\begin{equation}\label{eqn:dyn_system}
\frac{dx_\alpha(t)}{dt} = f_\alpha(\vec{x}(t)) + g_\alpha(\vec{u}(t)),
\end{equation}
where $\vec{x}$ with components $x_\alpha$ (where $\alpha=1,\dots, N_d$)
is the vector of observable state variables, $\vec{u}(t) \in \mathbb{R}^{N_c}$ is the motor command input, and $\vec{f}$ and $\vec{g}$ are vectors whose components are arbitrary non-linear functions $f_\alpha$ and $g_\alpha$ respectively.

\paragraph{}
Parameters of the spiking model neurons vary between one neuron and the next, both
in the command representation layer and the recurrent network, which yields  different frequency-current curves for different neurons (Supplementary Fig. \ref{fig:tuning_curves}).
Since arbitrary low-dimensional functions can be approximated by linear decoding from
a basis of nonlinear functions, such as neuronal tuning curves
\citep{funahashi_approximate_1989,girosi_networks_1990,hornik_multilayer_1989,sanner_gaussian_1992,eliasmith_neural_2004},
we may expect that a suitable  synaptic plasticity rule can tune the feedforward weights onto, and  the lateral weights within, the recurrent network so as to approximate the role of the functions $\vec{g}$ and $\vec{f}$ in equation \eqref{eqn:dyn_system}, respectively, while the read-out weights are kept fixed.

\paragraph{}
To enable weight tuning, we make four assumptions.
First, we assume that, during the learning phase, a random time-dependent motor command input $\vec{u}(t)$ is given to both the muscle-body reference system described by equation \ref{eqn:dyn_system}
and to the spiking network. The random input produces trajectories in the observable state variable corresponding to motor babbling \citep{petersson_spontaneous_2003}.
Second, we assume that each component  $\hat{x}_\alpha$ of the output predicted by the spiking network is compared to the actual observable output produced by the reference system of equation \ref{eqn:dyn_system} and their difference $\epsilon_\alpha$ is calculated, similar to supervised learning schemes such as perceptron learning \citep{rosenblatt_principles_1961}.
Third, we assume that this difference $\epsilon_\alpha = x_\alpha - \hat{x}_\alpha$ is projected back to neurons in the recurrent network through fixed random feedback weights with a large gain.
More precisely, our third assumption is that neuron $i$ receives a total error input $I^\epsilon_i = k \sum_\alpha e_{i\alpha} \epsilon_\alpha$ with feedback weights $k e_{i\alpha}$, where $k$ is fixed at a large constant value.
Fourth, we assume that the readout weights $d_{\alpha i}$ have been pre-learned by standard learning schemes \citep{voegtlin_temporal_2006,burbank_mirrored_2015}, so as to form an auto-encoder loop of gain $k$ with the fixed random feedback weights $ke_{i\alpha}$, i.e. an arbitrary value $\epsilon_\alpha$ sent via the error feedback weights to the recurrent network and read out, from its $N$ neurons, via the decoding weights gives back (approximately) $k\epsilon_\alpha$. From adaptive control theory \citep{narendra_stable_1989,ioannou_robust_2012} we may expect that the negative feedback arising from assumptions three and four, drives the neurons to generate a coarse activity pattern that leads to a close-to-ideal observable output (small errors), even at the very start of learning. Note that a vanishing error ($\epsilon_\alpha=0$ for all components), after a sufficiently long learning time, indicates that the neuronal network has autonomously generated the desired output so that feedback is no longer required.

\paragraph{}
While error feedback is on, the change in synaptic weights $w^{\textnormal{ff}}_{il}$ and $w_{ij}$ on the feedforward and recurrent connections, respectively, is:
\begin{align}
 \label{eqn:rec_error_learning}
 \dot{w}^{\textnormal{ff}}_{il}
 &= \eta \, (I_i^\epsilon * \kappa^\epsilon) (S^{\textnormal{ff}}_l*\kappa)(t), \nonumber \\
\dot{w}_{ij} &= \eta \,( I_i^\epsilon * \kappa^\epsilon ) (S_j*\kappa)(t),
\end{align}
where $\eta$ is the learning rate, and $\kappa^\epsilon$ is an exponentially decaying filter kernel with a time constant of 200 ms. For a postsynaptic neuron $i$, the error term $I^\epsilon_i * \kappa^\epsilon$ is the same for all its synapses, while the presynaptic contribution is synapse-specific. We call the learning scheme `Feedback-based Online Local Learning Of Weights' (FOLLOW), since the predicted state $\vec{\hat{x}}$ \textit{follows} the true state $\vec{x}$ from the start of learning. Under precise mathematical conditions, the FOLLOW scheme converges to a stable solution (Methods subsections \ref{lols_proof} and \ref{convergence_proof}).

\paragraph{}
We emphasize that the learning rule of equation \eqref{eqn:rec_error_learning} uses an error $\epsilon_\alpha \equiv x_\alpha - \hat{x}_\alpha$ in the observable state rather than an error in the derivative that would appear if descending down the gradient of a loss function (see Supplementary subsection \ref{rec_algo}) \citep{eliasmith_unified_2005,macneil_fine-tuning_2011}.
Furthermore, it is a local learning rule  since all quantities needed on the right-hand-side of equation \ref{eqn:rec_error_learning} could be available at the location of the synapse in the postsynaptic neuron.
For a potential local implementation scheme, let us imagine that the postsynaptic error current $I_i^\epsilon$  arrives in the apical dendrite where it stimulates  messenger molecules that quickly diffuse or are actively transported into the soma and  basal dendrites where synapses from feedfoward and feedback input could be located, as depicted in Figure \ref{fig:schematics}C. Consistent with the picture of a messenger molecule, we low-pass filter the error current with an exponential filter $\kappa^\epsilon$ of time constant 200 ms, much longer than the synaptic time constant of 20 ms of the filter $\kappa$. Simultaneously, filtered information about presynaptic spike arrival $S_j*\kappa$ is available at each synapse, possibly
in the form of glutamate bound to the postsynaptic receptor or by calcium triggered signaling chains localized in the postsynaptic spines. Thus the combination of effects caused by presynaptic spike arrival and error information available in the postsynaptic cell drives  weight changes, in loose analogy to standard Hebbian learning. For a critical evaluation of the notion of `local rule', we refer to the Discussion.

\subsection*{Spiking networks learn target dynamics via FOLLOW learning}

\paragraph{}
In order to check whether the FOLLOW scheme would enable the network to learn various dynamical systems, we studied three systems describing a nonlinear oscillator, low-dimensional chaos and simulated arm movements (additional examples in Supplementary Figures \ref{fig:lin}, \ref{fig:ffnonlin} and Methods). In all simulations, we started with vanishingly small feedforward and recurrent weights (tabula rasa), but assumed pre-learned readout weights matched to the error feedback weights. 
For each of the three dynamical systems, we had a learning phase and a testing phase. During each phase, we provided time-varying input to both the network (Fig. \ref{fig:schematics}B) and the reference system.
During the learning phase, rapidly changing control signals mimicked spontaneous movements (motor babbling) while synaptic weights were
updated according to the FOLLOW learning rule \eqref{eqn:rec_error_learning}.

\paragraph{}
During learning, the mean squared error, where the mean was taken over the number of dynamical dimensions $N_d$ and over a duration of a few seconds, decreased (Fig. \ref{fig:nonlin_spikes}D).
We stopped the learning phase i.e. weight updating, when the mean squared error approximately plateaued as a function of learning time (Fig. \ref{fig:nonlin_spikes}D).
At the end of the learning phase, we switched the error feedback off (`open loop') and provided different test inputs to both the reference system and the recurrent spiking network.
A successful forward predictive model should be able to predict the state variables in the open-loop model over a finite time horizon (corresponding to the planning horizon of a short action sequence) and in the closed-loop mode (with error feedback) without time limit.

\subsubsection*{Non-linear oscillator}
\label{vanderpol}
\begin{figure}
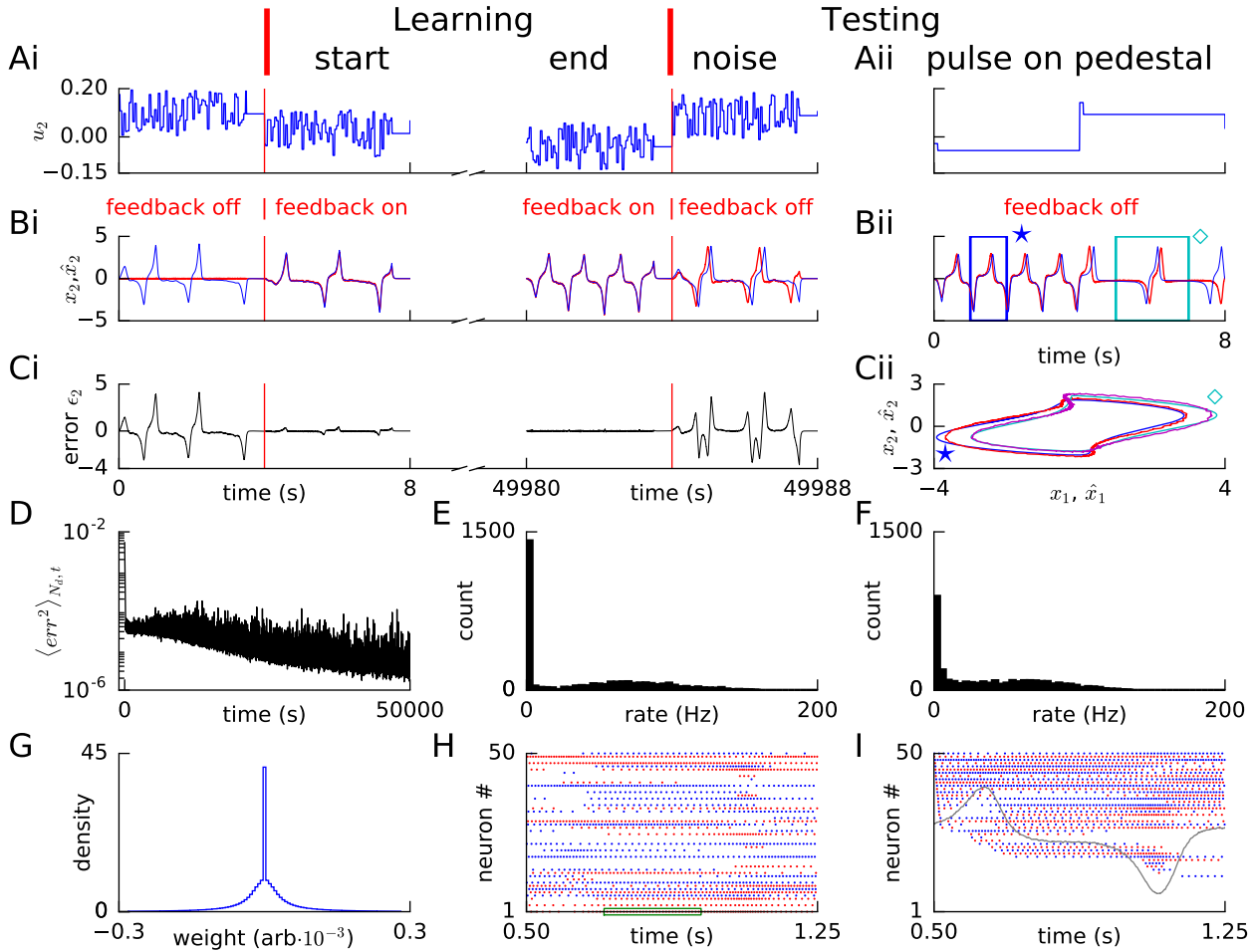

\centering
\includegraphics[width=\textwidth]{{{fig4_nonlin_altSpikes}}}
\caption{\textbf{Learning non-linear dynamics via FOLLOW: the van der Pol oscillator.}
\textbf{A-C.} Control input, output, and error, before and at the start of learning; in the last 4 s of learning; and during testing without error feedback (demarcated by the vertical red lines). Weight updating and error current feedback were both turned on after the vertical red line on the left at the start of learning, and turned off after the vertical red line in the middle at the end of learning. \textbf{A.} Second component of the input $u_2$. \textbf{B.} Second component of the learned dynamical variable $\hat{x}_2$ (red) decoded from the network, and the reference $x_2$ (blue). After the feedback was turned on, the output tracked the reference. The output continued to track the reference, even after the end of the learning phase, when feedback and learning were turned off. The output tracked the reference even with a very different input (Bii). \textbf{C.} Second component of the error $\epsilon_2=x_2-\hat{x}_2$ between the reference and the output. \textbf{Cii} Trajectory $(x_1(t),x_2(t))$ is the phase plane for reference (red,magenta) and prediction (blue,cyan) during two different intervals as indicated by $\star$ and $\diamond$ in Bii. \textbf{D.} Mean squared error per dimension averaged over 4 s blocks, on a log scale, during learning with feedback on. \textbf{E.} Histogram of firing rates of neurons in the recurrent network averaged over 0.25 s (interval marked in green in H) when output was fairly constant (mean across neurons was 36.95 Hz). \textbf{F.} As in E, but averaged over 16 s (mean across neurons was 37.36 Hz). \textbf{G.} Histogram of weights after learning. A few strong weights $|w_{ij}|>0.3$ are out of bounds and not shown here. \textbf{H.} Spike trains of 50 randomly-chosen neurons in the recurrent network (alternating colors for guidance of eye only). \textbf{I.} Spike trains of H, reverse-sorted by first spike time after 0.5 s, with output component $\hat{x}_2$ overlaid for timing comparison.}
\label{fig:nonlin_spikes}
\end{figure}

\paragraph{}
Our FOLLOW learning scheme enabled a network with 3000 neurons in the recurrent network and 3000 neurons in the motor command representation layer to approximate the non-linear 2-dimensional van der Pol oscillator (Fig. \ref{fig:nonlin_spikes}).
We used a superposition of random steps as input, with amplitudes drawn uniformly from an interval, changing on two time scales 50 ms and 4 s (see Methods).

\paragraph{}
During the 4 seconds before learning started, we blocked error feedback. Because of our initialization with zero feedforward and recurrent weights, the output $\hat{x}$ decoded from the network of spiking neurons remained constant at zero while the reference system performed the desired oscillations.
Once the error feedback with large gain was turned on, the feedback forced the network to roughly follow the reference.  Thus, with feedback, the error dropped to a very low value, immediately after the start of learning (Fig. \ref{fig:nonlin_spikes}B,C).
During learning, the error dropped even further over time (Fig. \ref{fig:nonlin_spikes}D).
After having stopped learning at 50000 s, we found the weight distribution to be uni-modal with a few very large weights (Fig. \ref{fig:nonlin_spikes}G). In the open-loop testing phase without error feedback, a sharp square pulse as initial input on different 4 s long pedestal values caused the network to track the reference as shown in Figure \ref{fig:nonlin_spikes}Aii-Cii panels. For some values of the constant pedestal input, the phase of the output of the recurrent network differed from that of the reference (Fig. \ref{fig:nonlin_spikes}Bii), but the shape of the non-linear oscillation was well predicted as indicated by the similarity of the trajectories in state space (Fig. \ref{fig:nonlin_spikes}Cii).

\paragraph{}
The spiking pattern of neurons of the recurrent network changed as a function of time, with inter-spike intervals of individual neurons correlated with the output, and varying over time (Fig. \ref{fig:nonlin_spikes}H,I). The distributions of firing rates averaged over a 250 ms period with fairly constant output, and over a 16 s period with time-varying output, were long-tailed, with the mean across neurons maintained at approximately 37 Hz (Fig. \ref{fig:nonlin_spikes}E,F). The latter distribution had lesser number of very low- and very high-firing neurons compared to the former, consistent with the expectation that the identity of low-firing and high-firing neurons changed over time for time-varying output (Fig. \ref{fig:nonlin_spikes}E,F). We repeated this example experiment (`van der Pol oscillator') with a network of equal size but with neurons that had firing rates increased by a factor of 2, so that some neurons could reach a maximal rate of 400 Hz (Supplementary Fig. \ref{fig:tuning_curves}). We found that the learning time was reduced by around 5 times (Supplementary Fig. \ref{fig:nonlin}) but with qualitatively the same behaviour. Hence, for all further simulations, we set neuronal parameters to enable firing rates up to 400 Hz (Supplementary Fig. \ref{fig:tuning_curves}B).

\subsubsection*{Chaotic Lorenz system}
\begin{figure}
\centering
\includegraphics[width=\textwidth]{{{fig5_lorenz}}}
\caption{\textbf{Learning chaotic dynamics via FOLLOW: the Lorenz system.}\newline
Layout and legend of panels \textbf{A-C} are analogous to Figure \ref{fig:nonlin_spikes}A-C. \textbf{D.} The trajectories of the reference (left panel) and the learned network (right panel) are shown in state space for 40 s with zero input during the testing phase, forming the well-known Lorenz attractor. \textbf{E.} Tent map, i.e. local maximum of the third component of the reference signal (blue) / network output (red) is plotted versus the previous local maximum, for 800 s of testing with zero input. The reference is plotted with filtering in panels A-C, but unfiltered for the strange attractor (panel D left) and the tent map (panel E blue).}
\label{fig:chaos}
\end{figure}

\paragraph{}
Our FOLLOW scheme also enabled a network with 5000 neurons each in the command representation layer and recurrent network, to learn the 3-dimensional non-linear chaotic Lorenz system (Fig. \ref{fig:chaos}). We considered a paradigm where the command input remained zero so that the network had to learn the autonomous dynamics characterized in chaos theory as a 'strange attractor' \citep{lorenz_deterministic_1963}.
During the testing phase without error feedback  minor differences led to different trajectories of the network and the reference which show up as large fluctuations of $\epsilon_3(t)$ (Fig. \ref{fig:chaos}A-C). Such a behavior is to be expected for a chaotic system where small changes in initial condition can lead to large changes in the trajectory.
Importantly, however, the activity of the spiking network exhibits qualitively the same underlying strange attractor dynamics, as seen from the butterfly shape \citep{lorenz_deterministic_1963} of the attractor in configuration space, and the tent map \citep{lorenz_deterministic_1963} of successive maxima versus the previous maxima (Fig. \ref{fig:chaos}D,E). The tent map generated from our network dynamics (Fig. \ref{fig:chaos}E) has lower values for the larger maxima compared to the reference tent map. However, very large outliers like those seen in a network trained by FORCE \citep{thalmeier_learning_2016} are absent. Since we expected that the observed differences are due to the filtering of the reference by an exponentially-decaying filter, we repeated learning without filtering the Lorenz reference signal (Supplementary Fig. \ref{fig:chaos_filtered}), and found that the mismatch is reduced, but a doubling appeared in the tent map which had been almost imperceptible with filtering (Fig. \ref{fig:chaos}E).

\subsection*{FOLLOW enables learning a two-link planar arm model under gravity}
\paragraph{}
To turn to a task closer to real life, we next wondered if a spiking network can also learn the dynamics of a two-link arm via the FOLLOW scheme.
We used a two-link arm model adapted from \citep{li_optimal_2006} as our reference.
The two links in the model correspond to the upper and fore arm, with the elbow joint in between and the shoulder joint at the top. The arm moved in the vertical plane under gravity, while torques were applied directly at the two joints, so as to coarsely mimic the action of muscles. To avoid full rotations, the two joints  were constrained to vary in the range from $-90^\circ$ to $+90^\circ$ where the resting state is at $0^\circ$ (see Methods).

\paragraph{}
The dynamical system representing the arm is four-dimensional with the state variables being the two joint angles and two angular velocities. The network must integrate the torques to obtain the angular velocities which in turn must be integrated for the angles. Learning these dynamics is difficult due to these sequential integrations involving nonlinear functions of the state variables and the input.
Because of the coupling of angles, torques, and moments, the dynamics of the two-link arm belong to a more general class of nonlinear differential equations, rather than equation \eqref{eqn:dyn_system}.
In particular, the general equations cannot be approximated by the combined feedforward and recurrent network architecture used until now, as the control input  is not  simply added to, but needs to be combined with the recurrent term. Thus, we use a modified network architecture with only the recurrent network, as described in Supplementary Section \ref{general_dyn_system}, and schematized in Supplementary Figure \ref{fig:general_dyn_schematic}.

\begin{figure}
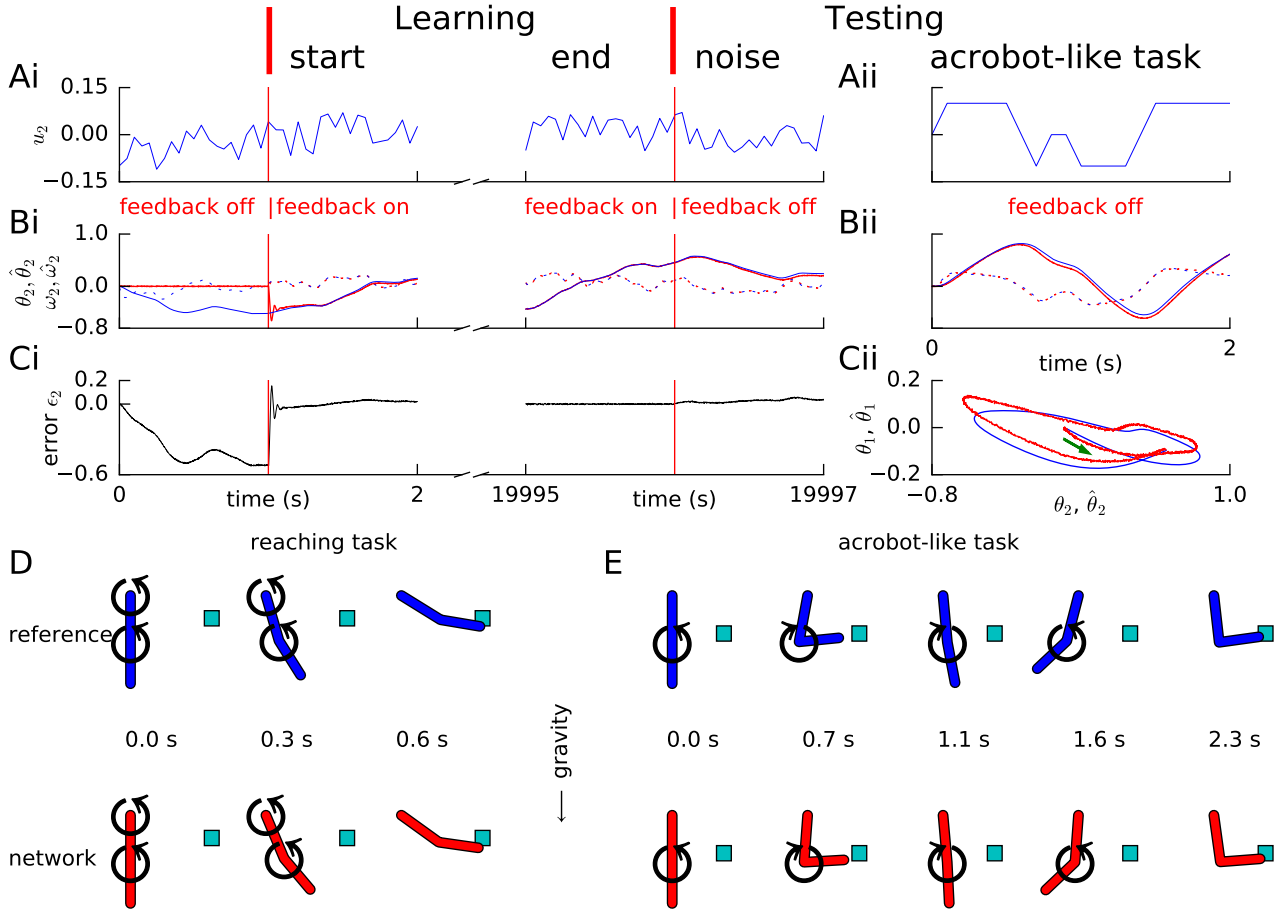

\centering
\includegraphics[width=\textwidth]{{{fig6_arm}}}
\caption{\textbf{Learning arm dynamics via FOLLOW.}
Layout and legend of panels \textbf{A-C} are analogous to Figure \ref{fig:nonlin_spikes}A-C except that: in panel \textbf{A}, the control input (torque) on the elbow joint is plotted; in panel \textbf{B}, reference and decoded angle $\theta_2,\hat{\theta}_2$ (solid) and angular velocity $\omega_2,\hat{\omega}_2$ (dotted) are plotted, for the elbow joint; in panel \textbf{C}, the error $\theta_2-\hat{\theta}_2$ in the elbow angle is plotted. \textbf{Aii-Cii.} The control input was chosen to perform a swinging acrobot-like task by applying small torque only on the elbow joint. \textbf{Cii.} The shoulder angle $\theta_1(t)$ is plotted versus the elbow angle $\theta_2(t)$ for the reference (blue) and the network (red) for the full duration in Aii-Bii. The green arrow shows the starting direction. \textbf{D.} Reaching task. Snapshots of the configuration of the arm, reference in blue (top panels) and network in red (bottom panels) subject to torques in the directions shown by the circular arrows. After 0.6 s, the tip of the forearm reaches the cyan target. Gravity acts downwards in the direction of the arrow. \textbf{E.} Acrobot-inspired swinging task (visualization of panels of Aii-Cii). Analogous to D, except that the torque is applied only at the elbow. To reach the target, the arm swings forward, back, and forward again.}
\label{fig:arm}
\end{figure}

\paragraph{}
Similar to the previous examples, random input torque with amplitudes of short and long pulses  changing each 50 ms and 1 s, respectively, was provided to each joint during the learning phase. The input was linearly interpolated between consecutive values drawn every 50 ms. In the closed loop scenario with error feedback, the trajectory converges rapidly to the target trajectory (Fig. \ref{fig:arm}). The general FOLLOW scheme learns to reproduce the arm dynamics even without error feedback for a few seconds during the test phase (Fig. \ref{fig:arm} and Supplementary Videos 1 and 2), which corresponds to the  time horizon needed for the planning of short arm movements.

\paragraph{}
To assess the generalization capacity of the network, we fixed the parameters postlearning, and tested the network in the open-loop setting on a reaching task and an acrobot-inspired swinging task \citep{sutton_generalization_1996}. In the reaching task, torque was provided to both joints to enable the arm-tip to reach beyond a specific $(x,y)$ position from rest.
The arm dynamics of the reference model and the network are illustrated in Figure \ref{fig:arm}D and animated in Supplementary Video 1. We also tested the learned network model of the 2-link arm on an acrobot-like task i.e. a gymnast swinging on a high-bar \citep{sutton_generalization_1996}, with the shoulder joint analogous to the hands on the bar, and the elbow joint to the hips. The gymnast can only apply small torques at the hip and none at the hands, and must reach beyond a specified $(x,y)$ position, by swinging. Thus, during the test, we provided input only at the elbow joint, with a time course that could make the reference reach beyond a specific $(x,y)$ position from rest by swinging. The control input and the dynamics (Figure \ref{fig:arm}A-C right panels, Figure \ref{fig:arm}E and Supplementary Video 2) show that the network can perform the task in open-loop condition suggesting that  it has learned the inertial properties of the arm model, necessary for this simplified acrobot task.

\subsection*{The FOLLOW scheme implicitly learns spike timings}
\begin{figure}
\centering
\includegraphics[width=\columnwidth]{{{fig7_error_to_zero_v2}}}
\caption{\textbf{Convergence of error, weights and spike times for a realizable reference network.} 
\newline 
\textbf{A.} We ran our FOLLOW scheme on a network for learning one of two different implementations of the reference van der Pol oscillator: (1) differential equations, versus (2) a network realized using FOLLOW learning for 10,000s. We plot the evolution of the mean squared error, mean over number of dimensions $N_d$ and over 4 s time blocks, from the start to 100,000 s of learning. With the weights starting from zero, mean squared error for the differential equations reference (1) is shown in black, while that for the realizable network reference (2) is in red. 
\textbf{B.} The feedforward weights (top panel) and the recurrent weights (bottom panel) at the end of 100,000s of learning, are plotted versus the corresponding weights of the realizable target network. The coefficient of determination i.e the $R^2$ value of the fit to the identity line ($y=x$) is also displayed for each panel. A value of $R^2 = 1$ denotes perfect equality of weights to those of the realizable network. Some weights fall outside the plot limits.
\textbf{C.} After 0 s, 10,000 s, and 100,000 s of the learning protocol against the realizable network as reference, we show spike trains of a few neurons in the recurrent network (red) and the reference network (blue) in the top, middle and bottom panels respectively, from test simulations while providing the same control input and keeping error feedback on.}
\label{fig:error_to_zero}
\end{figure}

\paragraph{}
In Methods subsections \ref{lols_proof} and \ref{convergence_proof}, we show that the FOLLOW learning scheme is Lyapunov stable and that the error tends to zero under certain reasonable assumptions and approximations.
Two important assumptions of the proof are that the weights remain bounded and that the desired dynamics are realizable by the network architecture, i.e. there exist feedforward and recurrent weights that enable the network to mimic the reference dynamics perfectly.
However, in practice the realizability is limited by at least two constraints.
First, even in networks of $N$ rate neurons with non-linear tuning curves, the nonlinear function $f$ of the reference system in equation \eqref{eqn:dyn_system} can in general only be approximated with a finite error \citep{funahashi_approximate_1989,girosi_networks_1990,hornik_multilayer_1989,sanner_gaussian_1992,eliasmith_neural_2004} which can be interpreted as a form of frozen noise, i.e. even with the best possible setting of the weights, the network predicts, for most values of the state variables, a next state which is slightly different than the one generated by the reference differential equation. Second, since we work with spiking neurons, we expect on top of this frozen noise the effect of shot noise caused by quasi-random spiking. Both noise sources may potentially cause drift of the weights \citep{narendra_stable_1989,ioannou_robust_2012} which in turn can make the weights leave the bounded regime. Ameliorative techniques from adaptive control are discussed in Supplementary subsection \ref{approx_error}. In our simulations, we did not find any effect of drift of weights on the error during a learning time up to 100,000 s (Fig. \ref{fig:error_to_zero}A), 10 times longer than that required for learning this example (Supplementary Fig. \ref{fig:nonlin}).

\paragraph{}
To highlight the difference between a realizable reference system and nonlinear differential equations as a reference system, we used a spiking network with fixed weights as the reference. For both the spiking reference network and the to-be-trained learning network we used the same  architecture, the same number of neurons, and the same neuronal parameters as in Supplementary Figure \ref{fig:nonlin} for the learning of the van der Pol oscillator.
However, instead of using the differential equations of the van der Pol oscillator as a reference, we now used as a reference the spiking network that was the final result after 10,000 s of FOLLOW learning in Supplementary Figure \ref{fig:nonlin}, i.e., a spiking approximation of the van der Pol oscillator. The read-out and feedback weights of the learning network had the same parameters as those of the spiking reference network, but the feedforward and recurrent weights of the learning network were initialized to zero and updated, during the learning phase, with the FOLLOW rule.
We ran FOLLOW learning against the reference network for 100,000 s (Supplementary Fig. \ref{fig:nonlin}).

\paragraph{}
We emphasize that, analogous to the earlier simulations, the feedback error
$\epsilon_\alpha$ was low-dimensional and calculated from the decoded outputs --- as opposed to supervised learning schemes for spiking neurons which are usually based on a direct comparison of spike times \citep{gutig_tempotron:_2006,pfister_optimal_2006,florian_chronotron:_2012,mohemmed_span:_2012,gutig_spike_2014,memmesheimer_learning_2014,gardner_supervised_2016}.
We found that, with the realizable network as a reference, learning was more rapid than with the original van der Pol oscillator as a reference. Interestingly, spike trains of the learning networks became similar to the spike trains in the reference network, even though matching of spike times was not used as an optimization criterion (Fig. \ref{fig:error_to_zero}C). In particular, a few neurons fired only two or three spikes at very precise moments in time. For example, after learning, the spikes of neuron $i=9$ in the learning network were tightly aligned with the spike times of the neuron with the same index $i$ in the spiking reference network. Similarly, neuron $i=8$ that was inactive at the beginning of learning  was found to be  active, and aligned with the spikes of the reference network, after 100,000 s of learning.

\paragraph{}
Moreover, network weights became very similar, though not completely identical, to the weights of the realizable reference network (Figure \ref{fig:error_to_zero}B), which suggests that the theorem for convergence of parameters from adaptive control  \citep{ioannou_robust_2012,narendra_stable_1989} should carry over to our learning scheme also.

\paragraph{}
Our results with the spiking reference network suggest that the error is reduced to a value close to zero for a realizable or closely-approximated system (Methods subsection \ref{convergence_proof}).

\subsection*{Learning is robust to sparse connectivity and noisy decoding weights}
\paragraph{}
So far, our spiking networks had all-to-all connectivity. We next tested whether sparse connectivity \citep{markram_reconstruction_2015,brown_intracortical_2009} of the feedforward and recurrent connections was sufficient for learning low-dimensional dynamics.
We ran the van der Pol oscillator learning protocol with the connectivity varying from 0.1 (10 percent connectivity) to 1 (full connectivity).
Connections that were absent after the sparse initialization could not appear during learning, while the existing sparse connections were allowed to evolve according to FOLLOW learning. As shown in Figure \ref{fig:randdec_sparsity}A, we found that learning was slower with sparser connectivity; but with twice the learning time, a sparse network with about 25\% connectivity reached similar performance  as the fully connected network with standard learning time.

\begin{figure}
\centering
\includegraphics[width=\columnwidth]{{{fig6_randdec_sparsity}}}
\caption{\textbf{Learning error versus sparse connectivity and noisy decoding weights.} 
\newline We ran the van der Pol oscillator learning protocol for 10,000 s for different parameter values and measured the mean squared error, over the last 400 s before the end of learning, mean over number of dimensions $N_d$ and time.
\textbf{A.} We evolved only a fraction of the feedforward and recurrent connections, randomly chosen as per a specific connectivity, according to FOLLOW learning, while keeping the rest zero. The round dots show the mean squared error for different connectivity after a 10,000 s learning protocol (default connectivity = 1 is starred); while the square dots show the same after a 20,000 s protocol.
\textbf{B.} We multiplied the original decoding weights (that form an auto-encoder with the error encoders) by a random factor (1+uniform$(-\chi,\chi)$) drawn for each weight. The mean squared error at the end of a 10,000s learning protocol for increasing values of $\chi$ is plotted (default $\chi = 0$ is starred).
\textbf{C.} We multiplied the original decoding weights by a random factor (1+uniform$(-\chi+\xi,\chi+\xi)$), fixing $\chi=2$, drawn for each weight. The mean squared error at the end of a 10,000 s learning protocol, for a few values of $\xi$ on either side of zero, is plotted.}
\label{fig:randdec_sparsity}
\end{figure} 

\label{random_decoders}
\paragraph{}
The read-out weights have been pre-learned until now, so that, in the absence of recurrent connections, error feedback weights and decoding weights formed an auto-encoder. We sought to relax this requirement.
Simulations showed that with completely random read-out weights, the system did not learn to reproduce the target dynamical system. However, if the read-out weights had some overlap with the auto-encoder, learning was still possible (Figure \ref{fig:randdec_sparsity}C). If for a feedback error $\vec{\epsilon}$, the error encoding followed by output decoding yields $k(1+\xi)\vec{\epsilon}+n(\vec{\epsilon})$, where $n$ is an arbitrary function, and $\xi$ is sufficiently greater than $-1$ so that the effective gain $k(1+\xi)$ remains large enough, then the term linear in error can still drive the output close to the desired one (see Methods).

\paragraph{}
To check this intuition in simulations, we incorporated multiplicative noise on the decoders by multiplying each decoding weight of the auto-encoder by one plus $\gamma$, where for each weight $\gamma$ was drawn independently from a uniform distribution between $-\chi+\xi$ and $\chi+\xi$. We found that the system was still able to learn the van der Pol oscillator up to $\chi \sim 5$ and $\xi=0$, or $\chi=2$ and $\xi$ variable (Figure \ref{fig:randdec_sparsity}B,C). Negative values of $\xi$ result in a lower overlap with the auto-encoder leading to the asymmetry seen in Figure \ref{fig:randdec_sparsity}C.

\paragraph{}
In conclusion, the FOLLOW learning scheme is robust to multiplicative noise on the decoding weights. Alternatively, decoder noise can also be studied as frozen noise in approximating the reference, causing a drift of the learned weights, possibly controlled by weight decay (Supplementary subsection \ref{approx_error}).

\section{Discussion}
\paragraph{}
The FOLLOW learning scheme enables a spiking neural network to function as a forward predictive model that mimics a non-linear dynamical system activated by one or several time-varying inputs. The learning rule is supervised, local, and comes with a proof of stability.

\paragraph{}
It is supervised because the FOLLOW learning scheme uses error feedback where the error is defined as the difference between predicted output and the actual observed output.
Error feedback forces the output of the system to mimic the reference, an effect that is widely used in adaptive control theory \citep{narendra_stable_1989,ioannou_robust_2012}.

\paragraph{}
The learning rule is local in the sense that it combines information about presynaptic spike arrival with an abstract quantity that we imagine to be available in the postsynaptic neuron. In contrast to standard Hebbian learning, the variable representing this postsynaptic quantity is not the postsynaptic firing rate, spike time, or postsynaptic membrane potential, but the error current projected by feedback connections onto the postsynaptic neuron, similar in spirit to modern biological implementation of approximated BackPropagation \citep{roelfsema_attention-gated_2005}, \citep{lillicrap_random_2016} or local versions of FORCE \citep{sussillo_generating_2009} learning rules.
We emphasize that the postsynaptic quantity is different from the postsynaptic membrane potential or the total postsynaptic current which would also include input from feedforward and recurrent connections. The separation of the  error current from the currents at  feedforward and recurrent synapses could be spatial (such as suggested in 
Fig. \ref{fig:schematics}C) or chemical if the error current projects onto synapses
that trigger a signaling cascade that is different from that at other synapses.
Importantly, whether it is a spatial or chemical separation, the signals triggered by the error currents need to be available throughout the postsynaptic neuron.

\paragraph{}
The learning rule is provenly stable with errors converging asymptotically to zero under a few assumptions (Methods subsection \ref{lols_proof}). The first assumption is that error encoding feedback weights and output decoding read-out weights form an auto-encoder. This requirement can be met if both sets of weights are learned at an early developmental stage, e.g., using mirrored STDP \citep{burbank_mirrored_2015}. The second assumption is that the reference dynamics $f(\vec{x})$ is realizable. This requirement can be approximately met by having a recurrent network with a large number $N$ of neurons with different parameters \citep{eliasmith_neural_2004}. The third assumption is that the state variables $\vec{x}(t)$ are observable. While currently we calculate the feedback error directly from the state variables as a difference between reference and predicted state, we could soften this condition and calculate the difference
in a higher-dimensional space with variables $\vec{y}(t)$ as long as $\vec{y} = K(\vec{x})$ is an invertible function of $\vec{x}(t)$ (Supplementary section \ref{general_dyn_system}).
The fourth assumption is that the system dynamics be slower than synaptic dynamics.
Indeed, typical reaching movements extend over hundreds of millisconds or a few seconds whereas neuronal spike transmission delays and synaptic time constants can be as short as a few milliseconds. In our simulations, neuronal and synaptic time constants were set to 20 ms, yet the network dynamics evolved on the time scale of hundreds of milliseconds or a few seconds, even in the open-loop condition when error feedback was switched off (Figs. \ref{fig:nonlin_spikes} and \ref{fig:arm}).
The fifth assumption is that weights stay bounded. Indeed, in biology, synaptic weights should not grow indefinitly. Algorithmically, a weight decay term in the learning rule can suppress the growth of large weights (see also Supplementary subsection \ref{approx_error}), though we did not need to implement a weight decay term in our simulations.

\paragraph{}
Simulations with the FOLLOW learning scheme have demonstrated that strongly nonlinear dynamics can be learned in a recurrent spiking neural network using a local online learning rule that does not require rapid weight changes. Previous work has mainly focused on a limited subset of  these aspects. For example, Eliasmith and colleagues used a learning rule derived from stochastic gradient descent, in a network structure comprising  heterogeneous spiking neurons with error feedback \citep{macneil_fine-tuning_2011}, but did not demonstrate learning non-linear dynamics (Supplementary subsection \ref{rec_algo}).
The mathematical framework of Den\`eve and colleagues with error feedback in a homogeneous spiking network with fast and slow connections yields a rule similar to ours, but has been applied to linear dynamics only \citep{bourdoukan_enforcing_2015}. 
It will be interesting to see if their approach to learning the autoencoder and maintaining balance could be used in our heterogeneous network.

\paragraph{}
Reservoir computing models exploit recurrent networks of non-linear units in an activity regime close to chaos where temporal dynamics is rich  \citep{jaeger_echo_2001,maass_real-time_2002,legenstein_input_2003,maass_computational_2004,jaeger_harnessing_2004,joshi_movement_2005,legenstein_edge_2007}.
While typical applications of reservoir computing are concerned with tasks involving a small set of desired output trajectories (such as switches or oscillators), our FOLLOW learning enables a recurrent network with a single set of parameters to mimic a dynamical system over a broad range of time-dependent inputs with a large family of different trajectories in the output.

\paragraph{}
Whereas initial versions of reservoir computing focused on learning the readout weights, applications of FORCE learning to recurrent networks of rate units made it possible to also learn the recurrent weights  \citep{sussillo_generating_2009, sussillo_transferring_2012}. However, in the case of a multi-dimensional target, multi-dimensional errors were typically fed to distinct parts of the network, as opposed to the distributed encoding used in our network. Moreover, the time scale of plasticity in FORCE learning is faster than the time scale of the dynamical system which is unlikely to be consistent with biology. Modern applications of FORCE learning to spiking networks
\citep{depasquale_using_2016,thalmeier_learning_2016,nicola_supervised_2016} inherit these issues.

\paragraph{}
Adaptive control of non-linear systems using continuous rate neurons \citep{sanner_gaussian_1992,slotine_adaptive_1987,slotine_adaptive_1986} and or spiking neurons \citep{dewolf_spiking_2016} has primarily focused on learning the weights on a feedforward processing path, but not on the weights in a recurrent network (note that adaptive control systems incorporate an error feedback loop but in the current context, we distinguish between the error feedback loop and the recurrent network).
Recurrent networks of rate units have occasionally been used for control \citep{zerkaoui_stable_2009}, but trained either via real-time recurrent learning or the extended Kalman filter which are non-local in space, or via backpropagation through time which is offline \citep{pearlmutter_gradient_1995}. Optimal control methods \citep{hennequin_optimal_2014} or stochastic gradient descent \citep{song_training_2016} have also been applied in recurrent networks of neurons, but with limited biological  plausibility of the published learning rules. As an alternative to supervised schemes, biologically plausible forms of reward-modulated Hebbian rules on the output weights of a reservoir have been used to learn periodic pattern generation and abstract computations \citep{hoerzer_emergence_2014,legenstein_reward-modulated_2010}, but how such modulated Hebbian rules could be used in predicting nonlinear dynamics for time-dependent control input remains open.

\paragraph{}
We found that the FOLLOW learning scheme does not require full connectivity but
also works with biologically more plausible sparse connectivity. Furthermore, it is robust to multiplicative noise in the output decoders, analogous to recent results on approximate error backpropagation in artificial neural networks \citep{lillicrap_random_2016}. Since the low-dimensional output and all neural currents are spatially averaged over a large number of synaptically-filtered spike trains, neurons in the FOLLOW network do not necessarily need to fire at rates higher than the inverse of the synaptic time scale.

\paragraph{}
Our present implementation of the FOLLOW learning scheme in spiking neurons violates Dale's law because synapses originating from the same presynaptic neuron can have positive or negative weights, but in a different context extensions incorporating Dale's law have been suggested \citep{parisien_solving_2008}. Neurons in cortical networks are also seen to maintain a balance of excitatory and inhibitory incoming currents \citep{deneve_efficient_2016}. It would be interesting to investigate a more biologically plausible extension of FOLLOW learning that maintains Dale's law; works in the regime of excitatory-inhibitory balance, possibly using inhibitory plasticity \citep{vogels_inhibitory_2011}; pre-learns the autoencoder, potentially via mirrored STDP \citep{burbank_mirrored_2015}; and possibly implements spatial separation between different compartments \citep{urbanczik_learning_2014}. Further directions worth pursuing include learning multiple different dynamical transforms within one recurrent network, without interference; hierarchical learning with stacked recurrent layers; and learning the inverse model of motor control so as to generate the control input given a desired state trajectory.

\section{Methods}

\subsection{Simulation software}
All simulation scripts were written in python (\url{https://www.python.org/}) for the Nengo 2 simulator \citep{stewart_python_2009} (\url{http://www.nengo.ca/}) with minor custom modifications to support sparse weights. We ran the model using the Nengo GPU backend (\url{https://github.com/nengo/nengo\_ocl}) for speed. The script for plotting the figures was written in python using the matplotlib module (\url{http://matplotlib.org/}). These simulation and plotting scripts will be made available online at \url{https://github.com/adityagilra}, once this article is peer-reviewed and published.

\subsection{Network architecture and parameters}
\label{network_architecture}
\paragraph{}
For all numerical simulations, we used deterministic leaky integrate and fire (LIF) neurons \citep{eliasmith_neural_2004}.
The voltage $V_l$ of each LIF neuron indexed by $l$, was a low-pass filter of its current $J_l$: $$\tau_m \frac{dV_l}{dt}=-V_l + J_l,$$ with $\tau_m$ the membrane time constant, set at 20 ms. The neuron fired when the voltage $V_l$ crossed a threshold $\theta=1$ from below, after which the voltage was reset to zero for a refractory period $\tau_r$ of 2 ms. If the voltage went below zero, it was clipped to zero. The output spike train of neuron $l$ is denoted $S^{\textnormal{ff}}_l(t)$ in the command representation layer and $S_l(t)$ in the recurrent network. Mathematically, a spike train is a sequence of events, modeled as a sum of Dirac delta-functions. To get the synaptic input current, spike trains were filtered with an exponential kernel $\kappa(t) \equiv \exp(-t/\tau_s)/\tau_s$ with a time constant of $\tau_s=20$ ms.

\paragraph{}
The low-dimensional motor command input $\vec{u}$, with $N_c$ components $u_\alpha$
was projected onto a layer of command-representing neurons. The current into a neuron with index $l$ was
\begin{equation}
\label{eqn:encoding_current}
J^{\textnormal{ff}}_l = \nu^{\textnormal{ff}}_l \sum_\alpha e^{\textnormal{ff}}_{k\alpha}u_\alpha +b^{\textnormal{ff}}_l,
\end{equation}
where $e^{\textnormal{ff}}_{k\alpha}$ were fixed random weights, while $b^{\textnormal{ff}}_l$ and $\nu^{\textnormal{ff}}_l$ were neuron-specific constants for bias and gain respectively (see below). We use Greek letters for the indices of  low-dimensional variables and Latin letters for neuronal indices, with summations going over the full range of the indices. The number of neurons $N$ in the command-representation layer was much larger than the dimensionality of the input i.e $N\gg N_c$.

\paragraph{}
A recurrent network of another $N$ neurons received feedforward input from the command-representation neurons, input from within the recurrent network, and error feedback. The input current to a neuron with index $i$ in the recurrent network was
\begin{equation}\label{eqn:adaptive_current}
J_i = \left(
        \sum_l w^{\textnormal{ff}}_{il} (S^{\textnormal{ff}}_l*\kappa)(t) +
        \sum_j w_{ij} (S_j*\kappa)(t) +
        \sum_\alpha k e_{i\alpha} (\epsilon_\alpha*\kappa)(t)
        \right) \nu_i + b_i,
\end{equation}
where $w^{\textnormal{ff}}_{il}$ and $w_{ij}$ were the feedforward and recurrent weights, respectively, which were both subject to our synaptic learning rule, whereas $ke_{i\alpha}$ were fixed error feedback weights (see below). The spike trains travelling along the feedforward path $S_l^{\textnormal{ff}}$ and those within the recurrent network $S_j$ were both low-pass filtered (convolution denoted by $*$) at the synapses with an exponential filter $\kappa$ with a time constant of 20 ms. The parameters $b_i$ and $\nu_i$ were neuron specific constants for the bias and gain, respectively. The constant $k>0$ was the feedback gain, and $\epsilon_\alpha \equiv x_\alpha-\hat{x}_\alpha$ was the output error. Here too, the number of neurons $N$ in the recurrent network was much larger than the dimensionality $N_d$ of the represented variable $\hat{x}$ i.e $N\gg N_d$.

\paragraph{}
Fixed, random encoding weights $e^{\textnormal{ff}}_{k\alpha}$ and error feedback weights (without factor $k$) $e_{i\alpha}$ were uniformly chosen on an $N_d$-dimensional hypersphere (Nengo 2 default) of radius $1/R_1$ and $1/R_2$ respectively. The norm of the variables, say $y_\alpha$ represented in the command representation layer (subscript 1) or the recurrent network (subscript 2) could range from $(0,R_{1,2})$ for $|\sum_\alpha e_{i\alpha} y_\alpha| < 1$, where $R_1$ and $R_2$ (Table \ref{tab:net_params}) will be called the representation radii of the two ensembles of $N$ neurons each. Typically $R_1$ was smaller than (around 0.2 of) $R_2$ as the input from the command representation layer was integrated in the recurrent network.

\paragraph{}
The intercept of the transfer function $a_i\left(\sum_\alpha e_{i\alpha} y_\alpha \right)$, i.e. firing rate vs. input curve, for each neuron (Fig. \ref{fig:tuning_curves}), was chosen uniformly from $(-1,1)$; and its maximum firing rate for an input within the representation radius, i.e. for $\sum_\alpha e_{i\alpha}y_\alpha = 1$, was chosen uniformly from $(100,200)$ Hz for the simulations in Figure \ref{fig:nonlin_spikes} and from $(200,400)$ Hz for all other simulations. From this desired intercept and maximum rate for the transfer function of each LIF neuron, with input $z_i=\gamma_i \sum_\alpha e_{i\alpha}y_\alpha + b_i$, its bias $b_i$ and gain $\gamma_i$ were calculated from the static rate equation for the LIF neuron \begin{equation}
\label{static_lif_equation}
a_i = 1 / (\tau_r + \tau_m \ln(1-1/z_i)). \end{equation}
Ideally, these encoders, biases and gains would be learnt during development from the input statistics, but here we set them randomly.

\paragraph{}
Decoders for the output $\vec{\hat{x}}$, i.e. the linear readout weights $d_{\beta i}$ from the recurrently connected network, were computed algorithmically to form an auto-encoder with respect to error-feedback weights $e_{i\alpha}$ (factor $k$ on these weights gives us gain $k$). To do this, we randomly selected $N$ error vectors, that we used as training points for optimization, $\epsilon^{(p)}_\beta$ where $1\le p \le N$ is the label of the training sample and $\beta$ is the index of the vector component. Since the observable system is $N_d$ dimensional, we chose the training points randomly from
an $N_d$-dimensional hypersphere of radius $R_1$. We applied each of the error vectors statically as input for the error feedback connections and calculated the activity $a^{(p)}_i$ of neuron $i$ for error vector $p$ using the static equation \eqref{static_lif_equation}.
The decoders $d_{\beta i}$ acting on these activities should yield back the encoded points thus forming an auto-encoder. A squared-error loss function 
\begin{equation}
\label{eqn:auto-encoder_expansive}
\mathcal{L} = \sum_p^P \sum_\beta \left( \sum_i^N d_{\beta i} a^{(p)}_i - \epsilon^{(p)}_\beta \right)^2 
\end{equation}
with L2 regularization was used for this linear regression (default in Nengo 2) \citep{eliasmith_neural_2004}. Biologically plausible learning rules exist for auto-encoders \citep{burbank_mirrored_2015,voegtlin_temporal_2006}, but we simply calculated and set the decoding weights as if they had already been learned.

\paragraph{}
Classical three-layer (input-hidden-output-layer) auto-encoders come in two different flavours, viz. compressive or expansive, which have the dimensionality of the hidden layer smaller or larger respectively, than that of the input and output layers. Instead of a three-layer feedfoward network, our auto-encoder forms a loop from the neurons in the recurrent network via readout weights to the output and from there via error-encoding weights to the input. Since the auto-encoder is in the loop, we expect that it works both as a compressive one (defined in equation \eqref{eqn:highDautoencoder}: from neurons in the recurrent network over the output back to the neurons) and as an expansive one (from the output through the neurons in the recurrent network back to the output). Rather than constraining the low-dimensional input $\epsilon_\beta$ and output $\sum_i^N d_{\beta i} a^{(p)}_i$ to be equal (expansive auto-encoder), we can enforce the high dimensional input $\sum_\beta e_{i\beta} \epsilon_\beta$ and output $\sum_{i,\beta}^N e_{j\beta} d_{\beta i} a^{(p)}_i$ to be equal (compressive auto-encoder) for computing the decoders of the auto-encoder. Thus the squared-error loss becomes:
\begin{align}
\mathcal{L}' =& \sum_p^P \sum_j \left( \sum_\beta e_{j\beta} (\sum_i^N d_{\beta i} a^{(p)}_i - \epsilon^{(p)}_\beta) \right)^2 \nonumber \\
=& \sum_p^P \sum_j \left( \sum_\beta e_{j\beta} (\sum_i^N d_{\beta i} a^{(p)}_i - \epsilon^{(p)}_\beta) \right)\left( \sum_\gamma e_{j\gamma} (\sum_l^N d_{\gamma l} a^{(p)}_l - \epsilon^{(p)}_\gamma) \right) \nonumber \\
=& \sum_p^P \sum_j \sum_\beta e_{j\beta}^2 \left(\sum_i^N d_{\beta i} a^{(p)}_i - \epsilon^{(p)}_\beta \right)^2 \nonumber \\
&+ \sum_p^P \sum_j \left( \sum_{\beta,\gamma,\beta\ne\gamma} e_{j\beta}e_{j\gamma} (\sum_i^N d_{\beta i} a^{(p)}_i - \epsilon^{(p)}_\beta) (\sum_l^N d_{\gamma l} a^{(p)}_l - \epsilon^{(p)}_\gamma) \right) \nonumber \\
\approx& \sum_p^P \sum_\beta \left(\sum_i^N d_{\beta i} a^{(p)}_i - \epsilon^{(p)}_\beta \right)^2 \nonumber,
\end{align}
where in the approximation, we consider that the term involving $\sum_j \sum_{\beta,\gamma,\beta\ne\gamma} e_{j\beta}e_{j\gamma}$ tends to zero as $e_{j\beta}$ and $e_{j\gamma}$ are independent; and further $\sum_j e_{j\beta}^2 = 1$. Thus, this loss function is approximately the same as the squared-error loss function in equation \eqref{eqn:auto-encoder_expansive} used for the expansive auto-encoder, showing that for an auto-encoder embedded in a loop, the expansive and compressive descriptions are equivalent. We employed a large number of random low-dimensional inputs when constraining the expansive auto-encoder.

\paragraph{}
The FOLLOW learning rule i.e. equation \eqref{eqn:rec_error_learning} was applied on the feedforward and recurrent weights, namely $w^{\textnormal{ff}}_{ji}$ and $w_{ji}$. The error for our learning rule was the error $\epsilon_\beta=x_\beta-\hat{x}_\beta$ in the observable output $\vec{x}$, not the error in the desired function $\vec{f}(\vec{x})$. The observable reference state $\vec{x}$ was obtained by integrating the differential equations of the dynamical system. The synaptic time constant $\tau_s$ was 20 ms in all synapses, including that for calculating the error and for feeding the error back to the neurons. The error used for the weight update was filtered by a 200 ms decaying exponential.

\paragraph{}
The command input vector $\vec{u}(t)$ to the network was $N_c$-dimensional ($N_c=N_d$ for all our simulations) and time-varying. During the learning phase, input changed on two different time scales. The fast value of each command component was switched every 50 ms to a level $u'_\alpha$ chosen uniformly between $(-\zeta_1,\zeta_1)$
and this number was added to a more slowly changing input variable $\bar{u}_\alpha$
(called 'pedestal' in the main part of the paper) which changed with a period $T_{period}$ as indicated in each of the figures. Here $\bar{u}_\alpha$ is the component of a vector of length $\zeta_2$ with a randomly chosen direction. The value of component $\alpha$ of the command is then $u_\alpha = \bar{u}_\alpha + u'_\alpha$.
Parameter values for the network and input for each dynamical system are provided in Table \ref{tab:net_params}. Further details are noted in subsection \ref{example_params}.

\paragraph{}
During the testing phase without error feedback, the network reproduced the reference trajectory of the dynamical system for a few seconds, in response to the same kind of input as during learning. We also tested the network on a different input not used during learning as shown in Figures \ref{fig:nonlin_spikes} and \ref{fig:arm}.

\subsection{Equations and parameters for the example dynamical systems}
\label{example_params}
\paragraph{}
The equations and input modifications for each dynamical system are detailed below. Time derivatives are in units of $s^{-1}$.

\paragraph{}
The equations for a linear decaying oscillators system (Supplementary Fig. \ref{fig:lin}) were
\begin{align*}
\dot{x}_1 &= u_1/0.02 + (-0.2 x_1 - x_2) / 0.05 \\
\dot{x}_2 &= u_2/0.02 + (x_1 - 0.2 x_2)/0.05.
\end{align*}
For this linear dynamical system, we tested the learned network on a ramp for 2 s followed by a step to a constant non-zero value.  A ramp can be viewed as a preparatory input before initiating an oscillatory movement, in a similar spirit to that observed in (pre-)motor cortex \citep{churchland_neural_2012}. For such input too, the network tracked the reference for a few seconds (Fig. \ref{fig:lin}A-C).

\paragraph{}
The equations for the van der Pol oscillator system were
\begin{align*}
\dot{x}_1 &= u_1/0.02 + x_2/0.125 \\
\dot{x}_2 &= u_2/0.02 + \left(2 (1-x_1^2)x_2-x_1\right)/0.125.
\end{align*}
Each component of the pedestal input $\bar{u}_\alpha$ was scaled differently for the van der Pol oscillator as reported in Table \ref{tab:net_params}.

\paragraph{}
The equations for the chaotic Lorenz system were
\begin{align*}
\dot{x}_1 &= u_1/0.02 + 10 (x_2 - x_1) \\
\dot{x}_2 &= u_2/0.02 - x_1 x_3 - x_2 \\
\dot{x}_3 &= u_3/0.02 + x_1 x_2 - 8(x_3+28)/3.
\end{align*}
In our equations above, $x_3$ of the Lorenz equations was represented by
an output variable $\tilde{x_3}=   x_3-28$ so as to have observable variables that
vary around zero. This does not change the system dynamics, just its representation in the network.
For the Lorenz system, only a pulse at the start for 250 ms, chosen from a random direction of norm $\zeta_1$, was provided to set off the system, after which the system followed autonomous dynamics.

\paragraph{}
Our FOLLOW scheme also learned non-linear feedforward transforms, as demonstrated in Supplementary Figures \ref{fig:ffnonlin} and \ref{fig:ffnonlinweights}. For the non-linear feedforward case, we used the linear system as the reference, but with the input transformed nonlinearly by $g_\alpha(\vec{u}) = 10((u_\alpha/0.1)^3 - u_\alpha/0.4)$. Thus, the equations of the reference were:
\begin{align*}
\dot{x}_1 &= 10((u_1/0.1)^3 - u_1/0.4) + (-0.2 x_1 - x_2) / 0.05 \\
\dot{x}_2 &= 10((u_2/0.1)^3 - u_2/0.4) + (x_1 - 0.2 x_2)/0.05.
\end{align*}
The input to the network remained $\vec{u}$. Thus, the feedforward weights had to learn the non-linear transform $\vec{g}(\vec{u})$ while the recurrent weights learned the linear system.

\begin{table}
\begin{tabular}{llllll}
~ & Linear & van der Pol & Lorenz & Arm & Non-linear feedforward \\
Number of neurons/layer & 2000 & 3000 & 5000 & 5000 & 2000 \\
$T_{period}$ (s) & 2 & 4 & 20 & 2 & 2 \\
Representation radius $R_1$ & 0.2 & 0.2 & 6 & - & 0.2 \\
Representation radius $R_2$ & 1 & 5 & 30 & 1 & 1 \\
Learning pulse $\zeta_1$ & $R_1/6$ & $R_1/6, R_1/2$ & $R_2/10$ & $R_2/0.3$ & $R_1/0.6$ \\
Learning pedestal $\zeta_2$ & $R_2/16$ & $R_1/6$, $R_1/2$ & 0 & $R_2/0.3$ & $R_2/1.6$ \\
\end{tabular}
\caption{Network parameters for example systems}
\label{tab:net_params} 
\end{table}

\paragraph{}
In the example of learning arm dynamics, we used a two-link model for an arm moving in the vertical plane with damping under gravity (see for example \url{http://www.gribblelab.org/compneuro/5\_Computational\_Motor\_Control\_Dynamics.html} and \url{https://github.com/studywolf/control/tree/master/studywolf\_control/arms/two\_link}), with parameters from \citep{li_optimal_2006}.
The differential equations for the four state variables, namely the shoulder and elbow angles $\vec{\theta} =(\theta_1,\theta_2)^T$ and the angular velocities $\vec{\omega} =(\omega_1,\omega_2)^T$, given input torques $\vec{\tau}=(\tau_1,\tau_2)^T$ were:
\begin{align}
\dot{\vec{\theta}} &= \vec{\omega}\\
\dot{\vec{\omega}} &= M(\vec{\theta})^{-1}\left(\vec{\tau}-C(\vec{\theta},\vec{\omega})-B\vec{\omega}-gD(\vec{\theta})\right)
\end{align}
with
$$
M(\vec{\theta}) = \left( \begin{matrix}
d_1+2d_2\cos\theta_2 + m_1s_1^2 + m_2s_2^2 & d_3+d_2\cos\theta_2+m_2s_2^2 \\
d_3+d_2\cos\theta_2+m_2s_2^2 & d_3+m_2s_2^2
\end{matrix} \right)
$$
$$
C(\vec{\theta},\vec{\omega})=\left(\begin{matrix}
-\dot{\theta}_2(2\dot{\theta}_1+\dot{\theta}_2) \\
                              {\dot{\theta}_1}^2
                             \end{matrix}\right)d_2\sin\theta_2,
B=\left(\begin{matrix}
         b_{11} & b_{12} \\
         b_{21} & b_{22}
        \end{matrix}\right),
$$
$$
D(\vec{\theta})=\left(\begin{matrix}
                       (m_1s_1+m_2l_1)\sin\theta_1 + m_2s_2\sin(\theta_1+\theta_2)\\
                       m_2s_2\sin(\theta_1+\theta_2)
                      \end{matrix}\right),
$$
$$ d_1 = I_1 + I_2 + m_2 l_1^2, d_2 = m_2 l_1 s_2 , d_3 = I_2, $$
where $m_i$ is the mass, $l_i$ the length, $s_i$ the distance from the joint center to the center of the mass, and $I_i$ the moment of inertia, of link $i$; $M$ is the moment of inertia matrix; $C$ contains centripetal and Coriolis terms; $B$ is for joint damping; and $D$ contains the gravitational terms. Here, the state variable vector $\vec{x} = [\theta_1,\theta_2,\omega_1,\omega_2]$, but the effective torque $\tau$ was obtained from the input torque $\vec{u}$ as below.

\paragraph{}
To avoid any link from rotating full 360 degrees, we provided an effective torque $\tau_\alpha$ to the arm, by subtracting a term proportional to the input torque $u_\alpha$, if the angle crossed $\pm$90 degrees and $u_\alpha$ was in the same direction:
$$
\tau_\alpha = u_\alpha -
\begin{cases}
      u_\alpha \tilde{\sigma}(\theta_\alpha) & u_\alpha > 0 \\
      0 & u_\alpha = 0 \\
      u_\alpha \tilde{\sigma}(-\theta_\alpha) & u_\alpha < 0
\end{cases},
$$
where $\tilde{\sigma}(\theta)$ increases linearly from 0 to 1 as $\theta$ goes from $\pi/2$ to $3\pi/4$ as below:
$$\tilde{\sigma}(\theta) =
\begin{cases}
  0 & \theta \le \pi/2 \\
  (\theta-\pi/2)/(\pi/4) & 3\pi/4>\theta>\pi/2 \\
  1 & \theta \ge 3\pi/4
\end{cases}
$$

\paragraph{}
The parameter values were as per Model 1 of the human arm in section 3.1.1 of the PhD thesis of Li \citep{li_optimal_2006} from the Todorov lab; namely $m_1=1.4 ~\text{kg}$, $m_2=1.1 ~\text{kg}$, $l_1 = 0.3 ~\text{m}$, $l_2 = 0.33 ~\text{m}$, $s_1=0.11 ~\text{m}$, $s_2= 0.16 ~\text{m}$, $I_1=0.025 ~\text{kg m}^2$, $I_2=0.045 ~\text{kg m}^2$, and $b_{11} = b_{22} = 0.05$, $b_{12} = b_{21} = 0.025$. Acceleration due to gravity was set at $g=9.81 ~\text{m/s}^2$. For the arm, we did not filter the reference variables for calculating the error.

\paragraph{}
The input torque $\vec{u}(t)$ for learning the two-link arm was generated, not by switching the pulse and pedestal values sharply, every 50 ms and $T_{period}$ as for the others, but by linearly interpolating in-between to avoid oscillations from sharp transitions, due to the feedback loop in the input in the general network (Supplementary Fig. \ref{fig:general_dyn_schematic}).

\paragraph{}
The input torque $\vec{u}$ and the variables $\vec{\omega}$, $\vec{\theta}$ obtained on integrating the arm model above were scaled by $0.02$, $0.05$ and $1./2.5$ respectively, and then used as the reference for the spiking network. Effectively, we scaled the input torques to cover one-fifth of the representation radius, the angular velocities one-half, and the angles full, as each successive variable was the integral of the previous one.

\subsection{Derivation and proof of stability of the FOLLOW learning scheme}
\label{lols_proof}
\paragraph{}
We derive the FOLLOW learning rules, while simultaneously proving the stability of the scheme. We assume that:
(1) the  feedback $\{k e_{i\alpha}\}$ and readout weights $\{d_{\alpha j}\}$ form an auto-encoder with gain $k$; (2) given the gains and biases of the spiking LIF neurons, there exist feedforward and recurrent weights that make the network follow the reference dynamics perfectly (in practice, the dynamics is only approximately realizable by our network, see Supplementary subsection \ref{approx_error} for a discussion); (3) the state $\vec{x}$ of the dynamical system is observable;
(4) the intrinsic time scales of the reference dynamics are much larger than the synaptic time scale, the time scale of the error feedback loop, and the time scale of learning; (5) the feedforward and recurrent weights remain bounded; and (6) the input $\vec{u}$ and reference output $\vec{x}$ remain bounded.

\paragraph{}
The proof proceeds in three major steps: using the auto-encoder assumption to write the evolution equation of the low-dimensional output state variable in terms of the recurrent and feedforward weights; showing that output follows the reference the error feedback loop; and obtaining the evolution equation for the error and using it in the time-derivative of a Lyapunov function $V$, to show that $\dot{V}\le 0$ for uniform stability, similar to proofs in adaptive control theory \citep{narendra_stable_1989,ioannou_robust_2012}.

\subsubsection*{Role of network weights for low-dimensional output}
\paragraph{}
The filtered low-dimensional output of the recurrent network is given by
\begin{equation}
\hat{x}_\alpha = \sum_j d_{\alpha j}(S_j*\kappa)(t)
\label{eqn:decoded_output}
\end{equation}
where $d_{\alpha j}$ are the readout weights. Since $\kappa$ is an exponential filter with time constant $\tau_s$, equation \eqref{eqn:decoded_output} can also be written as
\begin{equation}
\label{eqn:decoding_output}
 \tau_s\dot{\hat{x}}_\alpha(t) = - \hat{x}_\alpha(t) + \sum_j d_{\alpha j} S_j(t),
\end{equation}
We convolve this equation with kernel $\kappa$, multiply by the error feedback weights, and sum over the output components $\alpha$
\begin{equation}
\label{eqn:encoded_low_pass_output}
\tau_s\sum_\alpha e_{i\alpha}(\dot{\hat{x}}_\alpha*\kappa)(t) = - \sum_\alpha e_{i\alpha} (\hat{x}_\alpha*\kappa)(t) + \sum_\alpha e_{i\alpha} \sum_j d_{\alpha j}(S_j*\kappa)(t).
\end{equation}
We would like to write this evolution equation in terms of the recurrent and feedforward weight in the networks.

\paragraph{}
To do this, we exploit assumptions (1) and (4). Having shown the equivalence of the compressive and expansive descriptions of our auto-encoder in the error-feedback loop (Methods subsection \ref{network_architecture}), we formulate our nonlinear auto-encoder as compressive, which starts with a high-dimensional set of inputs $I_j\equiv (J_j-b_j)/\nu_j$ (where $J_j$ is the current into neuron $j$ having gain $\nu_j$ and bias $b_j$, cf. equations \eqref{eqn:encoding_current} and \eqref{eqn:adaptive_current}); transforms these nonlinearly into filtered spike trains $S_j[I_j]*\kappa$; decodes these filtered spike trains into a low-dimensional representation $\vec{z}$ with components $z_\alpha = \sum_j d_{\alpha j} (S_j[I_j]*\kappa)$; and blows-up the dimensionality back to the original one, via weights $ke_{i\alpha}$, to get inputs:
\begin{equation}
    I'_i = \sum_\alpha k e_{i\alpha} z_\alpha
    = k \sum_\alpha \sum_j e_{i\alpha} d_{\alpha j} (S_j[I_j]*\kappa).
\end{equation}
The spike train $S_j$ is a functional of the input $I_j$. $S_j$ depends on neuron index $j$ as neurons have different gains $\nu_j$ and biases $b_j$. Using assumption (1) we expect that the final inputs $I'_i$ are approximately $k$ times the initial inputs $I_i$:
\begin{equation}
\label{eqn:highDautoencoder}
k \sum_\alpha \sum_j e_{i\alpha} d_{\alpha j} (S_j[I_j]*\kappa) \approx k I_i \,.
\end{equation}

\paragraph{}
Our assumption (4) says that the state variables of the reference dynamics change slowly compared to neuronal dynamics. Due to the spatial averaging (sum over $j$ in equation \eqref{eqn:highDautoencoder}) over a large number of neurons, individual neurons do not necessarily have to fire at a rate higher than the inverse of the synaptic time scale, while we can still assume that the total round trip input $I'_i$ on the left hand side of equation $\eqref{eqn:highDautoencoder}$ is varying only on the slow time scale. Therefore, we used firing rate equations to compute mean outputs given static input when pre-calculating the readout weights (Methods subsection \ref{network_architecture}).

\paragraph{}
Inserting the approximate equation \eqref{eqn:highDautoencoder} in equation \eqref{eqn:encoded_low_pass_output} we find
\begin{equation}
\label{eqn:feedback_low_pass_output}
\tau_s \sum_\alpha e_{i\alpha}(\dot{\hat{x}}_\alpha*\kappa)(t)
 \approx -\sum_\alpha e_{i\alpha}(\hat{x}_\alpha*\kappa)(t) + I_i(t) .
\end{equation}
We replace $I_i \equiv (J_i-b_i)/\nu_i$, using the current $J_i$ from equation \eqref{eqn:adaptive_current} for neuron $i$ of the recurrent network, to obtain
\begin{align}
\label{eqn:low_pass_output_with_weights}
\tau_s \sum_\alpha e_{i\alpha}(\dot{\hat{x}}_\alpha*\kappa)(t)
\approx  & -\sum_\alpha e_{i\alpha}(\hat{x}_\alpha*\kappa)(t) + \sum_j w_{ij} (S_j*\kappa)(t) \nonumber \\ & ~~~~~ + \sum_l w^{\textnormal{ff}}_{il} (S^{\textnormal{ff}}_l*\kappa)(t) + \sum_\alpha k e_{i\alpha} (\epsilon_\alpha*\kappa)(t).
\end{align}
Thus, the change of the low-dimensional output $\hat{x}_\alpha * \kappa$ depends on the network weights, which need to be learned. This finishes the first step of the proof.

\subsubsection*{Error-feedback loop ensures that output follows reference}
\paragraph{}
Because of assumption (2), we may assume that there exists a recurrent network of spiking neurons that represents the desired dynamics of equation \eqref{eqn:dyn_system}
without any error feedback. This second network serves as a target during learning and has variables and parameters indicated with an asterisk. In particular, the second network has feedforward weights $w^{\textnormal{ff*}}_{il}$ and recurrent weights $w^*_{ij}$.
We write an equation similar to equation \eqref{eqn:feedback_low_pass_output} for the output $x^*_\alpha$ of the target network:
\begin{align}
\label{eqn:ideal_low_pass_output}
\tau_s \sum_\alpha e_{i\alpha}(\dot{{x}}^*_\alpha*\kappa)(t) =& -\sum_\alpha e_{i\alpha}({x}^*_\alpha*\kappa)(t) + \sum_j w^*_{ij} (S^*_j*\kappa)(t) \nonumber \\
 & ~~~~~~+ \sum_l w^{\textnormal{ff*}}_{il} (S^{\textnormal{ff*}}_l*\kappa)(t),
\end{align}
where $(S^{\textnormal{ff*}}_l*\kappa)(t)$ and $(S^*_j*\kappa)(t)$ are defined as the filtered spike trains of neurons in the realizable target network.
We emphasize that this target network does not need error feedback because its output is, by definition, always correct. In fact, the readout from the spike trains $S_j^*$ gives the target output which we denote by $\vec{x}^*$.
The weights of the target network are constant and their actual values are unimportant. They are mere mathematical devices to demonstrate stable learning of the first network which has adaptable weights.
For the first network, we choose the same number of neurons and the same neuronal parameters as for the second network; moreover, the weights from the command input to the representation layer and the readout weights from the recurrent network to the output are identical for both networks. Thus, the only difference is that the feedforward and recurrent weights of the target network are realized, while for the first network they need to be learnt.

\paragraph{}
In view of potential generalization, we note that any nonlinear dynamical system
is {\em approximately} realizable due to the expansion in a high-dimensional non-linear basis that is effectively performed by the recurrent network (see Supplementary subsection \ref{decoding}). Approximative weights (close to the ideal ones) could in principle also be calculated algorithmically as in Supplementary subsection \ref{rec_algo}. In the following we exploit assumption (2) and assume that the dynamics is actually (and not only approximately) realized by the target network.

\paragraph{}
Our assumption (3) states that the output is observable. Therefore the error component $\epsilon_\alpha$ can be computed directly via a comparison of the true output $\vec{x}$ of the reference with the output $\vec{\hat{x}}$ of the network: $\epsilon_\alpha = x_\alpha - \hat{x}_\alpha.$ (In view of potential generalizations, we remark that the observable output need not be the state variables themselves, but could be a higher-dimensional nonlinear function of the state variables, as shown for the general scheme in Supplementary section \ref{general_dyn_system}.)

\paragraph{}
As the second step of the proof, we now show that the error feedback loop enables the first network to follow the target network under assumptions (4)-(6). More precisely, we want to show that $\sum_\alpha e_{i\alpha} \hat{x}_\alpha \approx \sum_\alpha e_{i\alpha} x^*_\alpha$ for each neuron index $i$. To do so, we use assumption (4) and exploit that (i) learning is slow compared to the network dynamics so the weights of the first network can be considered momentarily constant; (ii) the reference dynamics is slower than the synaptic and feedback loop time scales, so the reference output $x_\alpha$ can be assumed momentarily constant. Thus, we have a separation of time scales in equation \eqref{eqn:low_pass_output_with_weights}: for a given input (transmitted via the feedforward weights) and a given target value $x^*_\alpha$, the network dynamics settles on the fast time scale $\tau_s$ to a momentary fixed point $\hat{x}^\dagger$ which we find by setting the derivative on the left-hand side of equation \eqref{eqn:low_pass_output_with_weights} to zero:
$$0 = -\sum_\alpha e_{i\alpha}(\hat{x}^\dagger_\alpha *\kappa)(t) + \sum_j w_{ij} (S_j*\kappa)(t) + \sum_l w^{\textnormal{ff}}_{il} (S^{\textnormal{ff}}_l*\kappa)(t) + \sum_\alpha k e_{i\alpha} ((x^*_\alpha-\hat{x}^\dagger_\alpha)*\kappa)(t).$$
We rewrite this equation in the form
\begin{equation}
\sum_\alpha e_{i\alpha}(\hat{x}^\dagger_\alpha *\kappa)(t) = \dfrac{k}{k+1}\sum_\alpha  e_{i\alpha} (x^*_\alpha*\kappa)(t) + \dfrac{1}{k+1} \left(\sum_j w_{ij} (S_j*\kappa)(t) + \sum_l w^{\textnormal{ff}}_{il} (S^{\textnormal{ff}}_l*\kappa)(t)\right).
\end{equation}
We choose the feedback gain for the error much larger than 1 ($k \gg 1$), such that $k/(k+1)\approx 1$. Using our assumption (5) that the feedforward and recurrent weights are bounded, and since the filtered spike trains remain bounded due to refractory period, the term in parentheses multiplying $1/(k+1)$ remains bounded, by say $B_1$. Choosing $k\gg B_1$, the second term can be made negligible. Thus, to obtain $\hat{x}^\dagger_\alpha \approx x^*_\alpha$, we set $k\gg 1$ and $k \gg B_1$.

\paragraph{}
To show that the fixed point is stable at the fast synaptic time scale, we calculate the Jacobian $\mathcal{J}=[\mathcal{J}_{il}]$, for the dynamical system given by equation \eqref{eqn:low_pass_output_with_weights}. We introduce auxiliary variables $y_i\equiv\sum_\alpha e_{i\alpha} \hat{x}_\alpha$ to rewrite equation \eqref{eqn:low_pass_output_with_weights} with the new variables in the form $\dot{y}_i = F_i(\vec{y})$; and then we take derivative of its right hand side to obtain the elements of the Jacobian matrix at the fixed point $\sum_\alpha e_{i\alpha} \hat{x}^\dagger_\alpha$:
$$
\mathcal{J}_{il} \equiv \dfrac{\partial F_i(\vec{y})}{\partial y_l} = -(k+1)\delta_{il}\int_{-\infty}^t \kappa(\tau)d\tau
+ \dfrac{\partial \sum_j w_{ij} (S_j*\kappa)(t) }{\partial y_l} \bigg|_{y_i=\sum_\alpha e_{i\alpha} \hat{x}^\dagger_\alpha},
$$
where $\delta_{il}$ is the Kronecker delta function. We note that $\sum_j w_{ij} (S_j*\kappa)$ is a spatially and temporally averaged measure of the population activity in the network with appropriate weighting factors $w_{ij}$. We assume that the population activity varies smoothly with input, which is equivalent to requiring that on the time scale $\tau_s$, the network fires asynchronously, i.e. there are no precisely timed population spikes. Then we can take the second term to be bounded, by say $B_2$. If we make $k\gg B_2$, then the Jacobian matrix $\mathcal{J}$ has large negative values on the diagonal and negligible values off-diagonal. Effectively, the Jacobian has negative eigenvalues, rendering the momentary fixed point asymptotically stable.

\paragraph{}
Thus, we have shown that if the initial state of the first network is close to the initial state of the target network, e.g. both start from rest, then on the slow time scale of the system dynamics of the reference $\vec{x}^*$, the first network follows the target network at all times, $\sum_\alpha e_{i\alpha} \hat{x}_\alpha \approx \sum_\alpha e_{i\alpha} x^*_\alpha$. With these constraints on each neuron, and since the readout weights, error-encoding weights and neuronal parameters are the same for the first and second network, the actual filtered spike trains of the recurrent neurons in the first network will be approximately the same as those of the target network, so that $(S_i*\kappa)(t)$ can be used instead of $(S^*_i*\kappa)(t)$ in \eqref{eqn:ideal_low_pass_output}. Moreover, the filtered spike trains $(S^{\textnormal{ff}}_l*\kappa)(t)$ of the command representation layer in the first network are always the same as those in the target network, since they are driven by the same command input $\vec{u}$ and the command encoding weights are, by construction, the same for both networks.

\subsubsection*{Stability of learning via Lyapunov's method}
\paragraph{}
We now turn to the third step of the proof and consider the temporal evolution of the error $\epsilon_\alpha = x_\alpha - \hat{x}_\alpha$. We exploit that the network dynamics is realized by the target network and insert equations \eqref{eqn:low_pass_output_with_weights} and \eqref{eqn:ideal_low_pass_output} so as to find
\begin{align}\label{eqn:epsilon_evolution}
\begin{split}
-\tau_s\sum_\alpha e_{i\alpha}(\dot{\epsilon}_\alpha*\kappa)(t)
                & = \tau_s\sum_\alpha e_{i\alpha}((\dot{\hat{x}}_\alpha-\dot{x}_\alpha)*\kappa)(t) \\
                & \approx \tau_s\sum_\alpha e_{i\alpha}((\dot{\hat{x}}_\alpha-\dot{x}^*_\alpha)*\kappa)(t) \\
                & \approx \sum_j \left( w_{ij} - w^*_{ij} \right) (S_j*\kappa)(t) + \sum_l \left( w^{\textnormal{ff}}_{il} - w^{\textnormal{ff*}}_{il} \right) (S^{\textnormal{ff}}_l*\kappa)(t) \\
                & ~~~~+ (k+1) \sum_\alpha e_{i\alpha} (\epsilon_\alpha*\kappa)(t) \\
                & \equiv \sum_j \psi_{ij} (S_j*\kappa)(t) + \sum_l \phi_{il} (S^{\textnormal{ff}}_l*\kappa)(t) + (k+1) \sum_\alpha e_{i\alpha} (\epsilon_\alpha*\kappa)(t),
\end{split}
\end{align}
In the second line, we have replaced the reference output by the target network output; and in the third line we have used equations \eqref{eqn:low_pass_output_with_weights} and \eqref{eqn:ideal_low_pass_output}, and replaced the filtered spike trains of the target network by those of the first network, exploiting the insights from the previous paragraph. In the last line, we have introduced abbreviations $\psi_{ij} \equiv w_{ij} - w^*_{ij}$ and $\phi_{il} \equiv w^{\textnormal{ff}}_{il} - w^{\textnormal{ff*}}_{il}$.

\paragraph{}
In order to show that the absolute value of the error decreases over time with an appropriate learning rule, we consider the candidate Lyapunov function:
\begin{equation}
\label{eqn:lyapunov_function}
V(\tilde{\epsilon},\psi,\phi) = \frac{1}{2}\sum_i \tilde{\epsilon}_i^2 + \frac{1}{2} \frac{1}{\tilde{\eta}_1} \sum_{i,j} (\psi_{ij})^2 + \frac{1}{2} \frac{1}{\tilde{\eta}_2} \sum_{i,l} (\phi_{il})^2,
\end{equation}
where $\tilde{\epsilon}_i \equiv \tau_s\sum_\alpha e_{i \alpha}(\epsilon_\alpha*\kappa)$ and
$\tilde{\eta}_1,\tilde{\eta}_2>0$ are positive constants.
The Lyapunov function is positive semi-definite $V(\tilde{\epsilon},\psi,\phi) \ge 0$, with the equality to zero only at $(\tilde{\epsilon},\psi,\phi)=(0,0,0)$. It has continuous first-order partial derivatives.

\paragraph{}
Furthermore, $V$ is \textit{radially unbounded} since
$$V(\tilde{\epsilon},\psi,\phi)>|(\tilde{\epsilon},\psi,\phi)|^2/(4\max(1,\tilde{\eta}_1,\tilde{\eta}_2)),$$
and \textit{decrescent} since
$$V(\tilde{\epsilon},\psi,\phi)<|(\tilde{\epsilon},\psi,\phi)|^2/\min(1,\tilde{\eta}_1,\tilde{\eta}_2),$$
where $|(\tilde{\epsilon},\psi,\phi)|^2\equiv\sum_i (\tilde{\epsilon}_i)^2 + \sum_{i,j} (\psi_{ij})^2 + \sum_{i,k} (\phi_{il})^2$ and $\min / \max$ take the minimum / maximum of their respective arguments.

\paragraph{}
Using Lyapunov's direct method, we need to prove, apart from the above conditions,
the property $\dot{V} \le 0$ for uniform global stability (which implies that
bounded orbits remain bounded, so the error remains bounded); or the stronger property $\dot{V}<0$ for asymptotic global stability (see for example \citep{narendra_stable_1989,ioannou_robust_2012}). Taking the time derivative of $V$, and replacing $\dot{\tilde{\epsilon}}_i$ i.e. $\tau_s\sum_\alpha e_{i\alpha}(\dot{\epsilon}_\alpha*\kappa)$ from \eqref{eqn:epsilon_evolution}, we have:
\begin{align}
\begin{split}
\dot{V} &= \sum_i \tilde{\epsilon}_i \dot{\tilde{\epsilon}}_i
        + \frac{1}{\tilde{\eta}_1} \sum_{i,j} \psi_{ij} \dot{\psi}_{ij}
        + \frac{1}{\tilde{\eta}_2} \sum_{i,l} \phi_{il} \dot{\phi}_{il}\\
        &\approx -\sum_i \tilde{\epsilon}_i
        \left( \sum_j \psi_{ij} (S_j*\kappa)(t) + \sum_l \phi_{il} (S^{\textnormal{ff}}_l*\kappa)(t) + (k+1)\sum_\alpha e_{i\alpha} (\epsilon_\alpha*\kappa)(t) \right) \\
        & ~~~~+ \frac{1}{\tilde{\eta}_1} \sum_{i,j} \psi_{ij} \dot{\psi}_{ij} + \frac{1}{\tilde{\eta}_2} \sum_{i,l} \phi_{il} \dot{\phi}_{il} \\
        &= \sum_{i,j} \psi_{ij} \left( -\tilde{\epsilon}_i (S_j*\kappa)(t) + \frac{1}{\tilde{\eta}_1} \dot{\psi}_{ij} \right) \\
        & ~~~~+ \sum_{i,k} \phi_{il} \left( -\tilde{\epsilon}_i (S^{\textnormal{ff}}_l*\kappa)(t) + \frac{1}{\tilde{\eta}_2} \dot{\phi}_{il} \right)
        - (k+1) \sum_i \tilde{\epsilon}_i^2 / \tau_s.
\end{split}
\end{align}

\paragraph{}
If we pick as a learning rule
\begin{align}
\label{eqn:stability_rule}
\dot{\psi}_{ij} &= \tilde{\eta}_1 \tilde{\epsilon}_i (S_j*\kappa)(t) \nonumber \\
\dot{\phi}_{il} &= \tilde{\eta}_2 \tilde{\epsilon}_i (S^{\textnormal{ff}}_l*\kappa)(t),
\end{align}
then $$\dot{V} = -(k+1) \sum_i \tilde{\epsilon}_i^2/\tau_s \le 0$$ choosing $k>-1$, which is subsumed under $k \gg 1$ for the error feedback. The condition \eqref{eqn:stability_rule} with $\eta_1\equiv\tilde{\eta}_1\tau_s$ and $\eta_2\equiv\tilde{\eta}_2\tau_s$, and $\kappa$ replaced by a longer filtering kernel $\kappa^\epsilon$, is the learning rule used in the main text, equation \eqref{eqn:rec_error_learning}.

\paragraph{}
Thus, in the $(\tilde{\epsilon},\psi,\phi)$-system given by equations \eqref{eqn:epsilon_evolution} and \eqref{eqn:stability_rule}, we have proven the global uniform stability of the fixed point $(\tilde{\epsilon},\psi,\phi) = (0,0,0)$, which is effectively $(\epsilon,\psi,\phi) = (0,0,0)$, choosing $\eta_1,\eta_2>0$ and $k\gg \max(1,B_1,B_2)$, under assumptions (1)-(6).

\paragraph{}
This ends our proof. So far, we have shown that the system is Lyapunov stable i.e. bounded orbits remain bounded, and not asymptotically stable. Indeed, with bounded firing rates and fixed readout weights, the output will remain bounded, as will the error (for a bounded reference). However, here, we also derived the FOLLOW learning rule, and armed with the inequality for the time derivative of the Lyapunov function in terms of the error, we further show in the following Methods subsection \ref{convergence_proof} that the error $\vec{\epsilon}$ goes to zero asymptotically, so that even without error feedback, $\vec{\hat{x}}$ reproduces the dynamics of $\vec{x}$ after learning.

\paragraph{}
A major caveat of this proof is that under assumption (2) the dynamics are {\em realizable} by our network. In a real application this might not be the case.
Approximation errors arising from a mismatch between the best possible network
and the actual target dynamics are currently ignored. The adaptive control literatue has shown that errors in approximating the reference dynamics appear as frozen noise and can cause runaway drift of the parameters \citep{narendra_stable_1989,ioannou_robust_2012}. In our simulations with a large number of neurons, the approximations of a non-realizable reference dynamics (e.g., the Van der Pol oscillator) were sufficiently good, and thus the expected drift was possibly slow, and did not cause the error to rise during typical timescales of learning.
A second caveat is our assumption (5). While the input is under our control and can therefore be kept bounded, some additional bounding is needed to stop weights from drifting. Various techniques to address such model-approximation noise and bounding weights have been studied in the robust adaptive control literature (e.g. \citep{ioannou_robust_1986,slotine_adaptive_1986,narendra_stable_1989,ioannou_adaptive_2006,ioannou_robust_2012}). We discuss this issue and briefly mention some of these ameliorative techniques in Supplementary sub-section \ref{approx_error}.

\paragraph{}
To summarize, the FOLLOW learning rule \eqref{eqn:stability_rule} on the feedforward or recurrent weights has two terms:
(i) a filtered presynaptic firing trace $(S^{\textnormal{ff}}_l*\kappa)(t)$ or $(S_j*\kappa)(t)$ that is available locally at each synapse; and (ii) a projected filtered error $\sum_{\alpha} e_{i\alpha} (\epsilon_\alpha*\kappa)(t)$ used for all synapses in neuron $i$ that is available as a current in the postsynaptic neuron $i$ due to error feedback, see equation \eqref{eqn:adaptive_current}. Thus the learning rule can be classified as local. Moreover, it uses an error in the observable $\vec{x}$, not in its time-derivative. While we have focused on spiking networks, the learning scheme can be easily used for non-linear rate units by replacing the filtered spikes $(S_i*\kappa)(t)$ by the output of the rate units $r(t)$. The rate units must also low-pass filter their input with kernel $\kappa(t)$.
Our proof is valid for arbitrary dynamical transforms $\vec{f}(\vec{x}) + \vec{g}(\vec{u})$ as long as they are realizable in a network. The proof does not involve gradient descent but shows uniform global stability using Lyapunov's method.

\subsection{Proof of error tending to zero asymptotically}
\label{convergence_proof}
\paragraph{}
In section \ref{lols_proof} we showed uniform global stability using $\dot{V} = -(k+1) \sum_i (\tilde{\epsilon}_i)^2 \le 0$, with $k\gg \max(1,B_1,B_2)$ and $\tilde{\epsilon}_i \equiv \tau_s \sum_\alpha e_{j \alpha}(\epsilon_\alpha*\kappa)$. This only means that bounded errors remain bounded. Here, we show more importantly that the error tends to zero asymptotically with time. We adapt the proof in section 4.2 of \citep{ioannou_robust_2012}, to our spiking network.

\paragraph{}
Here, we want to invoke a special case of Barb\u{a}lat's lemma:
if $f,\dot{f}\in\mathcal{L}_\infty$ and $f\in\mathcal{L}_p$ for some $p\in[1,\infty)$, then $f(t)\to 0$ as $t\to \infty$. Recall the definitions: function $f\in\mathcal{L}_p$ when $||x||_p\equiv\left(\int_0^\infty|f(\tau)|^p d\tau\right)^{1/p}$ exists (is finite); and similarly function $f\in\mathcal{L}_\infty$ when $||x||_\infty\equiv\sup_{t\ge0}|f(\tau)|$ exists (is finite).

\paragraph{}
Since $V$ is positive semi-definite ($V \ge 0$) and is a non-increasing function of time ($\dot{V}\le0$), its $\lim_{t\to\infty}V=V_\infty$ exists and is finite. Using this, the following limit exists and is finite:
$$
\sum_i \int_0^\infty (\tilde{\epsilon}_i(\tau))^2 d\tau = \frac{-1}{k+1}\int_0^\infty \dot{V}(\tau)d\tau = \frac{1}{k+1}(V(0)-V_\infty).
$$

\paragraph{}
Since each term in the above sum $\sum_i$ is positive semi-definite, $\int_0^\infty (\tilde{\epsilon}_i(\tau))^2 d\tau$ also exists and is finite $~\forall i$, and thus $\tilde{\epsilon}_i \in \mathcal{L}_2 ~\forall i$.

\paragraph{}
To show that $ \tilde{\epsilon}_i,\dot{\tilde{\epsilon}}_i \in \mathcal{L}_\infty ~\forall i$, consider equation \eqref{eqn:epsilon_evolution}. First, using the assumptions that the input $\vec{u}(t)$ is bounded and the reference dynamics is stable, we have that the reference output $\vec{x}(t)$ is bounded. Since network output $\vec{\hat{x}}$ is also bounded due to saturation of firing rates (as are the filtered spike trains), the error (each component) is bounded i.e. $\tilde{\epsilon}_i \in \mathcal{L}_\infty ~\forall i$. If we also bound the weights from diverging during learning, then $\psi_{ij},\phi_{il} \in \mathcal{L}_\infty ~\forall i,j,k$. With these reasonable assumptions, all terms on the right hand side of the equation \eqref{eqn:epsilon_evolution} for $\dot{\tilde{\epsilon}}_i$ are bounded, hence $\dot{\tilde{\epsilon}}_i \in \mathcal{L}_\infty ~\forall i$.

\paragraph{}
Since $\tilde{\epsilon}_i \in \mathcal{L}_2 ~\forall i$ and $ \tilde{\epsilon_i},\dot{\tilde{\epsilon}}_i \in \mathcal{L}_\infty ~\forall i$, invoking Barb\u{a}lat's lemma as above, we have $\tilde{\epsilon}_i\to 0 ~\forall i$ as $t\to \infty$. We have shown that the error tends to zero asymptotically under assumptions (1)-(6). In practice, the error shows fluctuations on a short time scale while the mean error over a longer time scale reduces and then plateaus, possibly due to approximate realizability, imperfections in the error-feedback, and spiking shot noise (cf. Fig. \ref{fig:error_to_zero}).

\paragraph{}
We do not further require the convergence of parameters to ideal ones for our purpose, since the error tending to zero, i.e. network output matching reference, is functionally sufficient for the forward predictive model. In the adaptive control literature \citep{ioannou_robust_2012,narendra_stable_1989}, the parameters are shown to converge to ideal ones if input excitation is ``persistent'', loosely that it excites all modes of the system. It should be possible to adapt the proof to our spiking network, as suggested by simulations (cf. Fig. \ref{fig:error_to_zero}), but is not pursued here.


\section{Acknowledgements}
\paragraph{}
We thank Johanni Brea, Samuel Muscinelli and Laureline Logiaco for helpful discussions and comments on a previous version of the manuscript. We thank Chris Stock, Tilo Schwalger, Olivia Gozel, and Dane Corneil for comments on the manuscript. Financial support was provided by the European Research Council (Multirules, grant agreement no. 268 689) and by the Swiss National Science Foundation (grant agreement no. CRSII\_147636).

\section{References}
\renewcommand\refname{\vskip -1cm}

\bibliography{zotero_neuroscience}

\newpage
\section{Supplementary Material}
\subsection{Decoding}
\label{decoding}
\paragraph{}
Consider only the command representation layer without the subsequent recurrent network. Assume, following \citep{eliasmith_neural_2004}, we wish to decode an arbitrary output $\vec{v}(\vec{u})$ corresponding to the $\vec{u}$ encoded in the command representation layer, from the spike trains $S^{\textnormal{ff}}_l(t)$ of the neurons, by synaptically filtering and linearly weighting the trains with decoding weights  $d^{(\vec{v})}_{\alpha l}$:
\begin{equation}
\label{eqn:arb_output}
\hat{v}_\alpha(\vec{u}) = \sum_l d^{(\vec{v})}_{\alpha l} (S^{\textnormal{ff}}_l*\kappa)(t),
\end{equation}
where $*$ denotes convolution $ (S^{\textnormal{ff}}_l*\kappa)(t) \equiv \int_{-\infty}^t S^{\textnormal{ff}}_l(t')\kappa(t-t')dt' = \int_{0}^{\infty} S^{\textnormal{ff}}_l(t-t')\kappa(t')dt'$, and $\kappa(t) \equiv \exp(-t/\tau_s)/\tau_s$ is a normalized filtering kernel.

\paragraph{}
We can obtain the decoders $d^{(\vec{v})}_{\alpha i}$ by minimizing the loss function
\begin{equation}
\label{eqn:general_loss_function}
L = \left\langle \sum_\alpha \left( v_\alpha(\vec{u}) - \sum_l d^{(\vec{v})}_{\alpha l} \langle S^{\textnormal{ff}}_l*\kappa \rangle_t \right)^2 \right\rangle_{\vec{u}}
\end{equation}
with respect to the decoders. The average $\langle \cdot \rangle_{\vec{u}}$ over $\vec{u}$ guarantees that the same constant decoders are used over the whole range of constant inputs $\vec{u}$. The time average $\langle \cdot \rangle_{t}$ denotes an analytic rate computed for each constant input for a LIF neuron. Linear regression with a finite set of constant inputs $\vec{u}$ was used to obtain the decoders (see Methods). With these decoders, if the input $\vec{u}$ varies slowly compared to the synaptic time constant $\tau_s$, we have $\hat{v}_\alpha = \sum_l d^{(\vec{v})}_{\alpha l}(S^{\textnormal{ff}}_l*\kappa)(t) \approx v_\alpha(\vec{u})$.

\paragraph{}
Any function of the input $\vec{v}(\vec{u})$ can be approximated with appropriate linear  decoding weights $d^{(\vec{v})}_{\alpha l}$ from the high-dimensional basis of non-linear tuning curves of heterogeneous neurons with different biases, encoding weights and gains, schematized in Figure \ref{fig:schematics}B. With a large enough number of such neurons, the function is expected to be approximated to arbitrary accuracy.
While this has not been proven rigorously for spiking neurons, this has theoretical underpinnings from theorems on universal function approximation using non-linear basis functions \citep{funahashi_approximate_1989,hornik_multilayer_1989,girosi_networks_1990}
sucessful usage in spiking neural network models by various groups \citep{seung_stability_2000,eliasmith_neural_2004,eliasmith_unified_2005,pouget_spatial_1997}, and biological plausibility \citep{poggio_theory_1990,burnod_visuomotor_1992,pouget_spatial_1997}.

\paragraph{}
Here, the neurons that are active at any given time operate in the mean driven regime i.e. the instantaneous firing rate increases with the input current \citep{gerstner_neuronal_2014}. The dynamics is dominated by synaptic filtering, and the membrane time constant does not play a significant role \citep{eliasmith_neural_2004,eliasmith_unified_2005,seung_stability_2000,abbott_building_2016}. Thus, the decoding weights derived from equation \eqref{eqn:general_loss_function} with stationary input are good approximations even in the time-dependent case, as long as the input varies on a time scale slower than the synaptic time constant.

\subsection{Online learning based on a loss function and its shortcomings}
\label{rec_algo}

\paragraph{}
The dynamical system given by equation \eqref{eqn:dyn_system} is to be mimicked by our spiking network implementing a different dynamical system with an extra error feedback term as in equation \eqref{eqn:low_pass_output_with_weights}. This can be interpreted as:
\begin{equation}
\label{eqn:rec_adaptive_dynamics}
\tau_s \dot{\hat{x}}_\alpha = -\hat{x}_\alpha + \breve{f}_\alpha(\vec{\hat{x}}) + \breve{g}_\alpha(\vec{u}) + k \epsilon_\alpha.
\end{equation}
Comparing with the reference equation \eqref{eqn:dyn_system}, after learning we want that $\breve{f}_\alpha(\vec{\hat{x}})$ and $\breve{g}_\alpha(\vec{u})$ should approximate $\tilde{f}_\alpha(\vec{\hat{x}}) \equiv \tau_s f_\alpha(\vec{\hat{x}})+\hat{x}_\alpha$ and $\tilde{g}_\alpha(\vec{u}) \equiv \tau_s g_\alpha(\vec{u})$ respectively (upto a constant term distributed between the two functions). In our simulations, we will usually start with zero feedforward and  recurrent weights, so that initially $\breve{f}(\vec{\hat{x}})=0=\breve{g}_\alpha(\vec{u})$.]

\paragraph{}
Assuming that the time scales of dynamics are slower than synaptic time scale $\tau_s$, we can approximate the requisite feedforward and recurrent weights, by minimizing the following loss functions respectively, with respect to the weights \citep{eliasmith_neural_2004}:

\begin{equation}\label{eqn:ff_loss_function}
L_\textnormal{ff} = \left\langle \sum_j \left(\sum_\alpha e^{\textnormal{ff}}_{k\alpha} \tilde{g}_\alpha(\vec{u}) - \sum_l w_{jl} \langle S_l^{\textnormal{ff}}*\kappa \rangle_t \right)^2 \right\rangle_x,
\end{equation}

\begin{equation}\label{eqn:rec_loss_function}
L_{rec} = \left\langle \sum_j \left(\sum_\alpha e_{j\alpha} \tilde{f}_\alpha(\vec{x}) - \sum_i w_{ji} \langle S_i*\kappa \rangle_t \right)^2 \right\rangle_x.
\end{equation}

\paragraph{}
Using these loss functions, we can precalculate the weights required for any dynamical system numerically, similarly to the calculation of decoders in Supplementary subsection \ref{decoding}.

\paragraph{}
We now derive rules for learning the weights online based on stochastic gradient descent of these loss functions, similar to \citep{macneil_fine-tuning_2011}, and point out some shortcomings.

\paragraph{}
The learning rule for the recurrent weights by gradient descent on the loss function given by equation \eqref{eqn:rec_loss_function} is
\begin{align}
\label{eqn:rec_loss_rule}
\begin{split}
\dfrac{d w_{ji} }{dt} & = -\dfrac{1}{2}\eta \dfrac{\partial L_{rec}}{\partial w_{ji}} \\
          & \approx \eta\left\langle \left(\sum_\beta e_{j\beta} \tilde{f}_\beta(\vec{x}) - \sum_i w_{ji} (S_i*\kappa)(t) \right) (S_i*\kappa)(t) \right\rangle_x \\
          & \equiv \eta\left\langle \epsilon^{(\tilde{f})}_j (S_i*\kappa)(t)\right\rangle_x. 
\end{split}
\end{align}
In the second line, the effect of the weight change on the filtered spike trains is assumed small and neglected, using a small learning rate $\eta$. With requisite dynamics slower than synaptic $\tau_s$, and with large enough number of neurons, we have approximated $ \sum_i w_{ji} \langle S_i*\kappa \rangle_t (t) \approx \sum_i w_{ji} (S_i*\kappa)(t)$. The third line defines an error in the projected $\vec{\tilde{f}}(\vec{x})$, which is the supervisory signal.

\paragraph{}
If we assume that the learning rate is slow, and the input samples the range of $x$ uniformly, then we can remove the averaging over $x$, similar to stochastic gradient descent.
\begin{equation}
\dfrac{d w_{ji}}{d t} \approx -\eta \epsilon^{(\tilde{f})}_j (S_i*\kappa)(t).
\end{equation}
where $\epsilon^{(\tilde{f})}_j \equiv \left(\sum_\beta e_{j\beta} \tilde{f}_\beta(\vec{x}) - \sum_i w_{ji} (S_i*\kappa)(t) \right)$.
This learning rule is the product of a projected multi-dimensional error $\epsilon^{(\tilde{f})}_j$ and the filtered presynaptic spike train $(S_i*\kappa)(t)$. However, this projected error in the unobservable $\vec{\tilde{f}}$ is not available to the postsynaptic neuron, making the learning rule non-local. A similar issue arises in the feedforward case.

\paragraph{}
In mimicking a dynamical system, we want only the observable output of the dynamical system i.e. $\vec{x}$ to be used in a supervisory signal, not a term involving the unknown $f(\vec{x})$ appearing in the derivative $\dot{\vec{x}}$. Even if this derivative is computed from the observable $\vec{x}$, it will be noisy. Furthermore, this derivative cannot be obtained by differentiating the observable versus time, if the observable is not directly the output, but an unknown non-linear function of it, which however our FOLLOW learning can handle (see Supplementary subsection \ref{general_dyn_system}). Thus, an online rule using just the observable error can learn only an integrator for which $f(x) \sim x$ \citep{macneil_fine-tuning_2011}.

\paragraph{}
Indeed learning both the feedforward and recurrent weights simultaneously using gradient descent on these loss functions, requires two different and unavailable error currents to be projected into the postsynaptic neuron to make the rule local.

\subsection{General dynamical system}
\label{general_dyn_system}
\paragraph{}
General dynamical systems of the form
\begin{align}
\frac{d\vec{x}(t)}{dt} &= \vec{h}(\vec{x}(t),\vec{u}(t)), \nonumber \\
\vec{y}(t) &= \vec{K}(\vec{x}(t)) \nonumber
\label{eqn:general_dyn_system}
\end{align}
can be learned using a different network configuration than the Figure \ref{fig:schematics}A configuration used for systems of the form \eqref{eqn:dyn_system}. Here, the state variable is $\vec{x}$, but the observable which serves as the reference to the network is $\vec{y}$. We handle this general system in two steps, first by absorbing the transformation equation of the observable (second equation) into the first equation, and then using an augmented variable and a different network structure.

\paragraph{}
Consider the transformation equation for the observable. The dimensionality of the relevant variables: (1) the state variables (say joint angles and velocities) $\vec{x}$; (2) the observables represented in the brain (say sensory representations of the joint angles and velocities) $\vec{y}$; and (3) the control input (motor command) $\vec{u}$, can be different from each other, but must be small compared to the number of neurons. Furthermore, we require the observable $\vec{y}$ to not lose information compared to $\vec{x}$, i.e. $\vec{K}$ must be invertible, so $\vec{y}$ will have at least the same dimension as $\vec{x}$.

\paragraph{}
The time evolution of the observable is
$$\dot{y}_\beta=\sum_\alpha \dfrac{\partial K_\beta (\vec{x})}{\partial x_\alpha} \dot{x}_\alpha = \sum_\alpha \dfrac{\partial K_\beta (\vec{x})}{\partial x_\alpha} h_\alpha(\vec{x},\vec{u}) \equiv p_\beta(\vec{y},\vec{u}).$$
The last step follows since function $\vec{K}$ is invertible, so that $\vec{x}=\vec{K}^{-1}(\vec{y})$. So we essentially need to learn $\dot{y}_\beta = p_\beta(\vec{y},\vec{u})$.

\paragraph{}
Having solved the observable transformation issue, we use $\vec{x}$ now for our observable, consistent with the main text. The dynamical system to be learned is now $\dot{x}_\beta = h_\beta(\vec{x},\vec{u})$. Consider an augmented vector of the state and input variables $\tilde{x}_\gamma \equiv [x_\alpha, u_\beta]$, where the index $\gamma$ runs over the values for indices $\alpha$ and $\beta$ serially. Its time derivative is $\dot{\tilde{x}}_\gamma = [\dot{x}_\alpha,\dot{u}_\beta] = [h_\alpha(\vec{\tilde{x}}),\dot{u}_\beta]$. The $\alpha$ components are of the same form as \eqref{eqn:dyn_system}, but the $\beta$ components involve $\dot{u}_\beta$ which is not specified as a function $h_\beta(\vec{\tilde{x}})$. However, we do not need $\dot{u}_\beta$ as only the state variables need to be predicted, not the input.

\paragraph{}
Our network for the general dynamical system, as shown in Supplementary Figure \ref{fig:general_dyn_schematic}, decodes the augmented $\vec{\hat{\tilde{x}}}$ i.e. $\vec{\hat{x}}$ and $\vec{\hat{u}}$ both as output. The error in the augmented $\vec{\tilde{x}}$ is fed back to the network. Since the $h_\alpha(\vec{\tilde{x}})$ is a mixed function of the state and input variables, $\vec{u}$ is no longer fed into the network via a command representation layer, rather it enters only via the error in the augmented variable. Once learning is complete, the error feedback in $\vec{\hat{x}}$ can be stopped, but the error in $\vec{\hat{u}}$ must still be fed back, so that $\vec{u}$ functions as a motor command.

\subsection{Approximation error causes drift in weights}
\label{approx_error}
\paragraph{}
A frozen noise term $\xi(\vec{x}(t))$ due to the approximate decoding from non-linear tuning curves of neurons, by the feedforward weights, recurrent weights and output decoders, will appear additionally in equation \eqref{eqn:epsilon_evolution}.
If this frozen noise has a non-zero mean over time as $\vec{x}(t)$ varies, leading to a non-zero mean error, then it causes a drift in the weights due to the error-based learning rules in equations \eqref{eqn:rec_error_learning}, and possibly a consequent increase in error. Note that the stability and error tending to zero proofs assume that this frozen noise is negligible.

\paragraph{}
Multiple strategies with contrasting pros and cons have been proposed to counteract this parameter drift in the robust adaptive control literature \citep{ioannou_robust_2012,narendra_stable_1989,ioannou_adaptive_2006}. These include a weight leakage / regularizer term switched slowly on, when a weight crosses a threshold \citep{ioannou_robust_1986,narendra_stable_1989}, or a dead zone strategy with no updation of weights once the error is lower than a set value \citep{slotine_adaptive_1986,ioannou_robust_2012}. In our simulations, the error continued to drop even over longer than typical learning time scales (Figure \ref{fig:error_to_zero}), and so, we did not implement these strategies.

\paragraph{}
In practice, the learning can be stopped once error is low enough, while the error feedback can be continued, so that the learned system does not deviate too much from the observed one.

\newpage
\renewcommand\thefigure{S\arabic{figure}}
\subsection{Supplementary Figures}
\setcounter{figure}{0}

\begin{figure}[H]
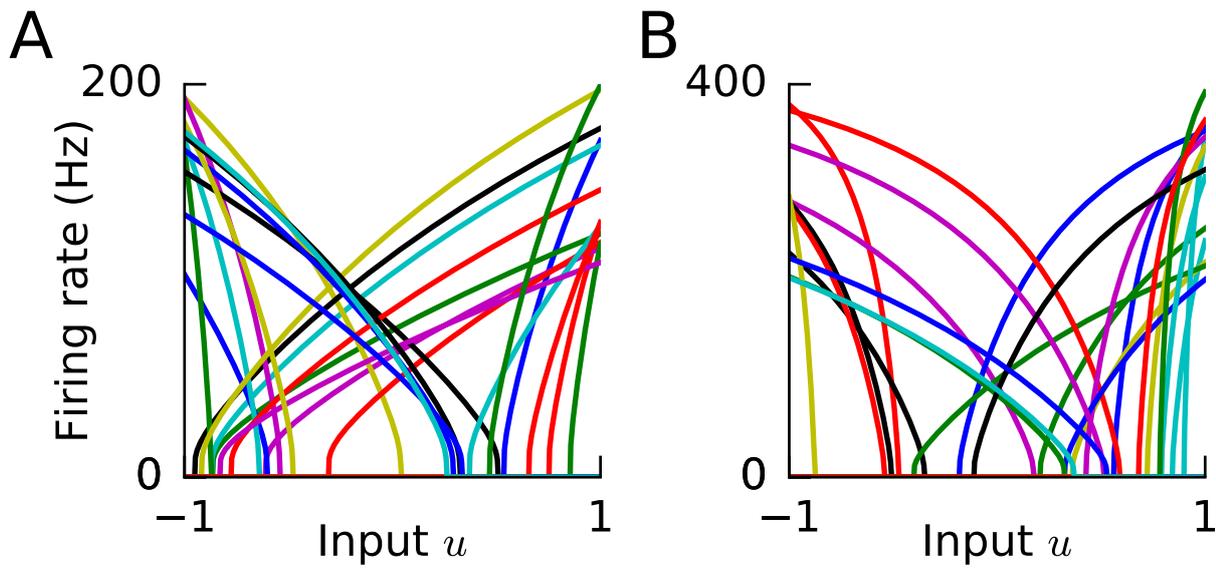

\centering
\includegraphics[width=\textwidth]{{{fig_tuning_curves}}}
\caption{\textbf{Tuning curves of heterogeneous neurons} The neurons in the command representation layer and the recurrent network were heterogeneous with different tuning curves (random gains/encoders/biases). \textbf{A,B.} Illustrative tuning curves of a few neurons, to projected input $u_i=\sum_\alpha e_{i\alpha} u_\alpha$ into neuron $i$, are plotted, with maximal firing rate of \textbf{A.} 200 Hz as for the network of Figure \ref{fig:nonlin_spikes}; and \textbf{B.} 400 Hz as for all other simulations (cf. Supplementary Figure \ref{fig:nonlin}).}
\label{fig:tuning_curves}
\end{figure}

\begin{figure}
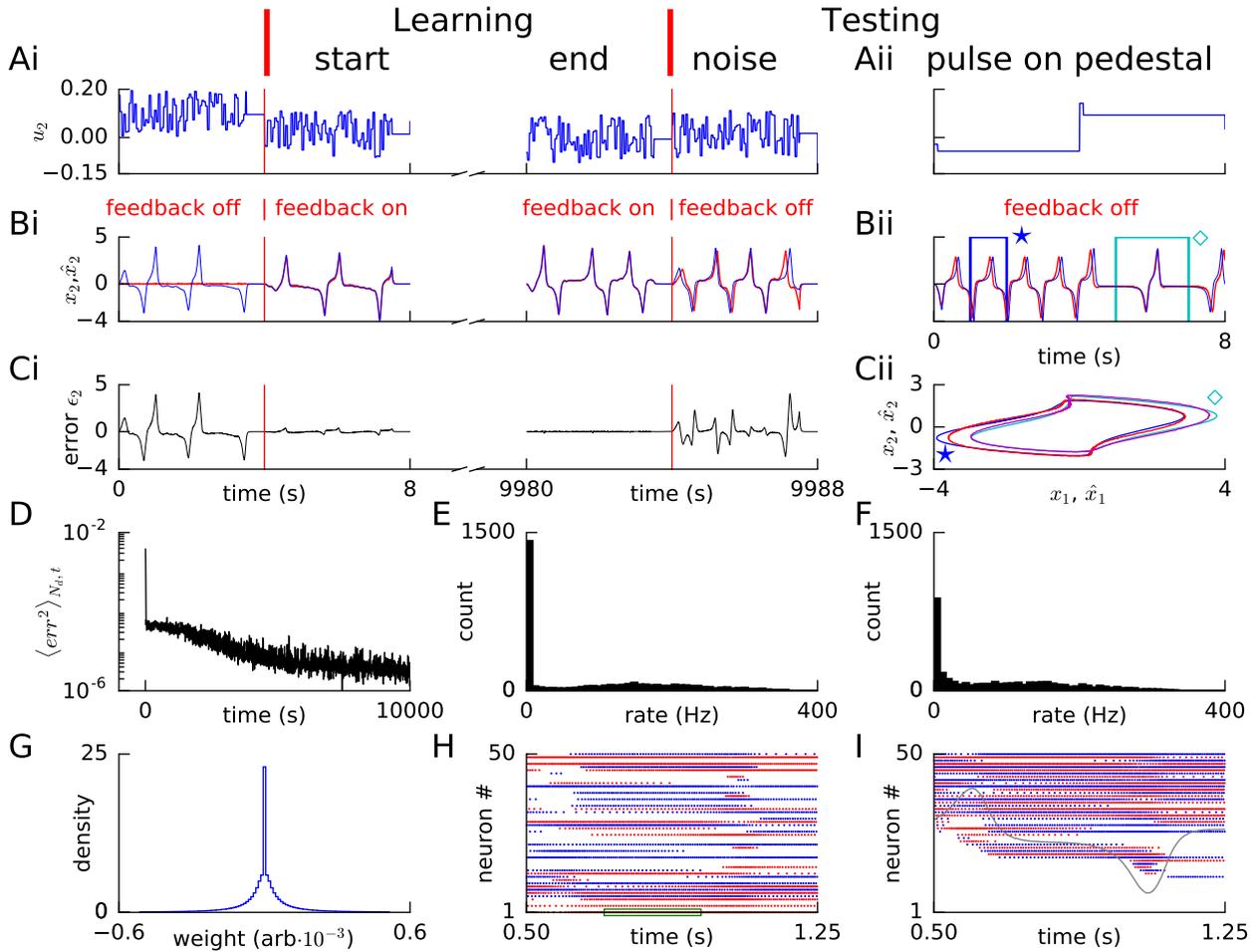

\centering
\includegraphics[width=\textwidth]{{{fig4_nonlin_highrates}}}
\caption{\textbf{Learning van der Pol oscillator dynamics via FOLLOW with double firing rates.} Layout and legend of panels \textbf{A-I} are analogous to Figure \ref{fig:nonlin_spikes}A-I, except that the gains of neurons were doubled yielding approximately double firing rates, but one-fifth the learning time.}
\label{fig:nonlin}
\end{figure}

\begin{figure}
\centering
\includegraphics[width=\textwidth]{{{fig_lorenz_norefinptau}}}
\caption{\textbf{Learning the Lorenz system without filtering the reference variables.}\newline
Panels \textbf{A-E} are interpreted similar to Figure \ref{fig:chaos}A-E, except that the reference signal was not filtered in computing the error. Without filtering, the tent map for the network output (panel E red) shows a doubling, but the mismatch for large maxima is reduced, compared to with filtering in Figure \ref{fig:chaos}.}
\label{fig:chaos_filtered}
\end{figure}

\begin{figure}
\centering
\includegraphics[width=0.67\columnwidth]{{{rec_error_general_schematic_v2}}}
\caption{\textbf{Network configuration for learning a general dynamical system} \newline
The recurrent network decodes the augmented vector $\hat{\tilde{x}}_\gamma = (\vec{\hat{x}},\vec{\hat{u}})$. Compare this configuration with that in Figure \ref{fig:schematics}A. The input is no longer provided directly to the network but only via the augmented error signal. The augmented error signal is fed back into the recurrent layer. Only the recurrent weights need to be learned, as there is no feedforward layer here. After learning is turned off, the error in input must continue to be fed-back to the neurons, to serve as a motor command; but the error in the state variables need not be.}
\label{fig:general_dyn_schematic}
\end{figure}

\begin{figure}
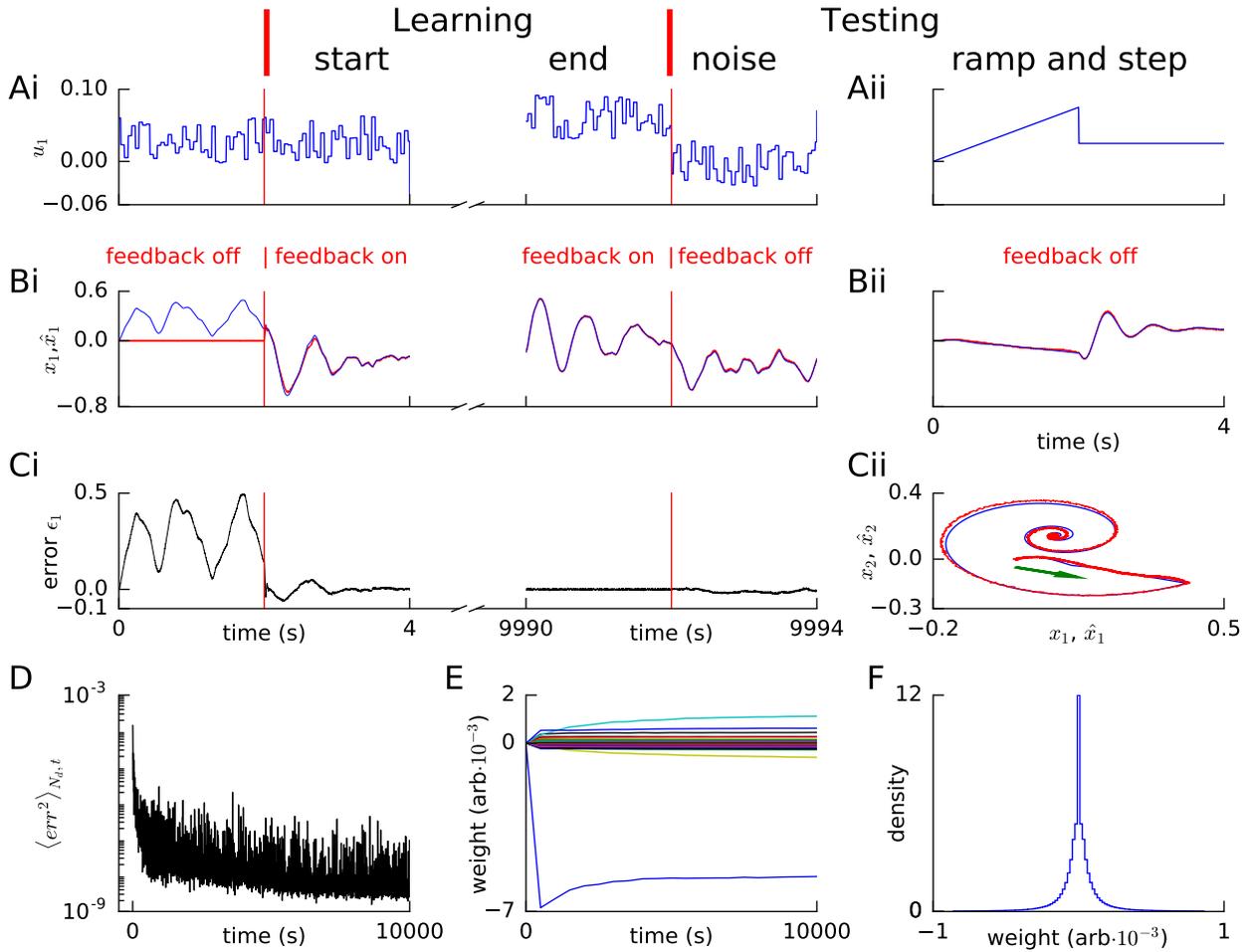

\centering
\includegraphics[width=1.0\textwidth]{{{fig3_lin}}}
\caption{\textbf{Learning linear dynamics via FOLLOW: 2D decaying oscillator.} Layout and legend of panels \textbf{A-D} are analogous to Figure \ref{fig:nonlin_spikes}A-D, except an error component is shown in the right-most panel of C. \textbf{E.} A few randomly selected weights are shown evolving during learning. \textbf{F.} Histogram of weights after learning. A few strong weights $|w_{ij}|>1$ are out of bounds and not shown here (cf. panel E).}
\label{fig:lin}
\end{figure}

\begin{figure}
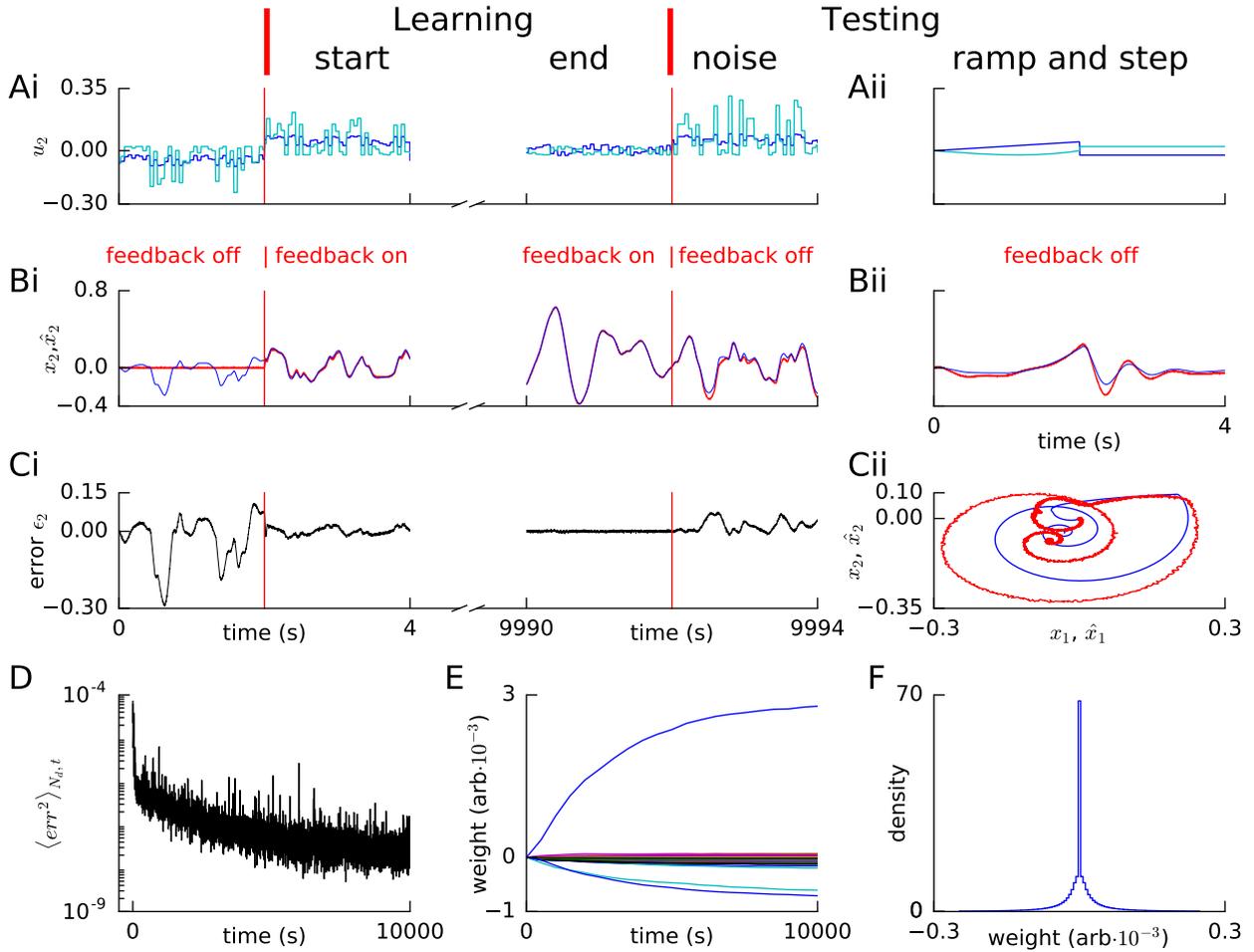

\centering
\includegraphics[width=\textwidth]{{{fig_lin_nonlin2}}}
\caption{\textbf{Learning nonlinear feedforward transformation with linear recurrent dynamics via FOLLOW.}
Panels \textbf{A-F} are interpreted similar to Figure \ref{fig:lin}A-F, except in panel A, along with the input $u_2(t)$ (blue) to layer 1, the required non-linear transform $g_2(\vec{u}(t))/20$ is also plotted (cyan); and in panels E and F, the evolution and histogram of weights are those of the feedforward weights, that perform the non-linear transform on the input.}
\label{fig:ffnonlin}
\end{figure}

\begin{figure}
\centering
\includegraphics[width=\columnwidth]{{{fig8_ff_nonlin_rec_compare}}}
\caption{\textbf{Feedforward weights are uncorrelated, while recurrent ones are correlated, when learning same recurrent dynamics but with different feedforward transforms.} \newline The linear decaying oscillators system was learned for 10,000s with either a non-linear or a linear feedforward transform. \textbf{A.} The learned feedforward weights $w^{\textnormal{ff}}_{ij}$ were plotted for the system with the non-linear feedforward transform, versus the corresponding feedforward weights for the system with the linear feedforward transform. The feedforward weights for the two systems do not fit the identity line (coefficient of determination $R^2$ is negative; $R^2$ is not the square of a number and can be negative) showing that the learned feedforward transform is different in the two systems as expected.
\textbf{B.} Same as \textbf{A}, but for the recurrent weights in the two systems. The weights fit the identity line with an $R^2$ close to 1 showing that the learned recurrent transform is similar in the two systems as expected. Some weights fall outside the plot limits.}
\label{fig:ffnonlinweights}
\end{figure} 

\end{document}